\documentclass{eptcs}

\usepackage[T1]{fontenc}
\usepackage[latin9]{inputenc}
\usepackage{array}
\usepackage{ifthen}
\usepackage{slashed}
\usepackage{amstext}
\usepackage{wasysym}
\usepackage{graphicx}
\usepackage{amssymb}

\makeatletter

\AtBeginDocument{}

\DeclareFontEncoding{LGR}{}{}

\newcommand{\lyxmathsym}[1]{\ifmmode\begingroup\def\b@ld{bold}
  \text{\ifx\math@version\b@ld\bfseries\fi#1}\endgroup\else#1\fi}

\title{Two-Domain DNA Strand Displacement}

\author{Luca Cardelli
\institute{Microsoft Research\\ Cambridge, UK}
\email{luca@microsoft.com}
}

\def\titlerunning{Two-Domain DNA Strand Displacement}
\def\authorrunning{Luca Cardelli}

\begin{document}
\maketitle

\begin{abstract}
We investigate the computing power of a restricted class of DNA strand displacement structures: those that are made of double strands with nicks (interruptions) in the top strand. To preserve this structural invariant, we impose restrictions on the single strands they interact with: we consider only two-domain single strands consisting of one toehold domain and one recognition domain. We study fork and join signal-processing gates based on these structures, and we show that these systems are amenable to formalization and to mechanical verification.
\end{abstract}

\section{Introduction }

Among the many techniques being developed for molecular computing
\cite{Hagiya. Towards Molecular Programming}, \textit{DNA strand
displacement} has been proposed as mechanism for performing computation
with DNA strands \cite{Seelig. Enzyme-Free Nucleic Acid Logic Circuits,Fontana. Pulling Strings}.
In most schemes, single-stranded DNA acts as \textit{signals} and
double-stranded (or more complex) DNA structures act as \textit{gates}.
Various circuits have been demonstrated experimentally \cite{Yurke. Using DNA to Power Nanostructures}.
The strand displacement mechanism is appealing because it is \textit{autonomous}
\cite{Green. DNA Hairpins: Fuel for Autonomous DNA Devices}: once
signals and gates are mixed together, computation proceeds on its
own without further intervention until the gates or signals are depleted
(output is often read by fluorescence). The energy for computation
is provided by the gate structures themselves, which are turned into
inactive waste in the process. Moreover, the mechanism requires \textit{only}
DNA molecules: no organic sources, enzymes, or transcription/translation
ingredients are required, and the whole apparatus can be chemically
synthesized and run in basic wet labs. 

The main aims of this approach are to harness computational mechanisms
that can operate at the molecular level and produce nano-scale structures
under program control, and somewhat separately that can intrinsically
interface to biological entities \cite{Benenson. Programmable and Autonomous Computing Machine made of Biomolecules}.
The computational structures that one may easily implement this way
(without some form of unbounded storage) vary from Boolean networks,
to state machines, to Petri nets. The last two are particularly interesting
because they take advantage of DNA's ability to encode symbolic information:
they operate on DNA strands that represent abstract signals. 

\begin{figure}
\begin{centering}
\includegraphics[scale=0.75]{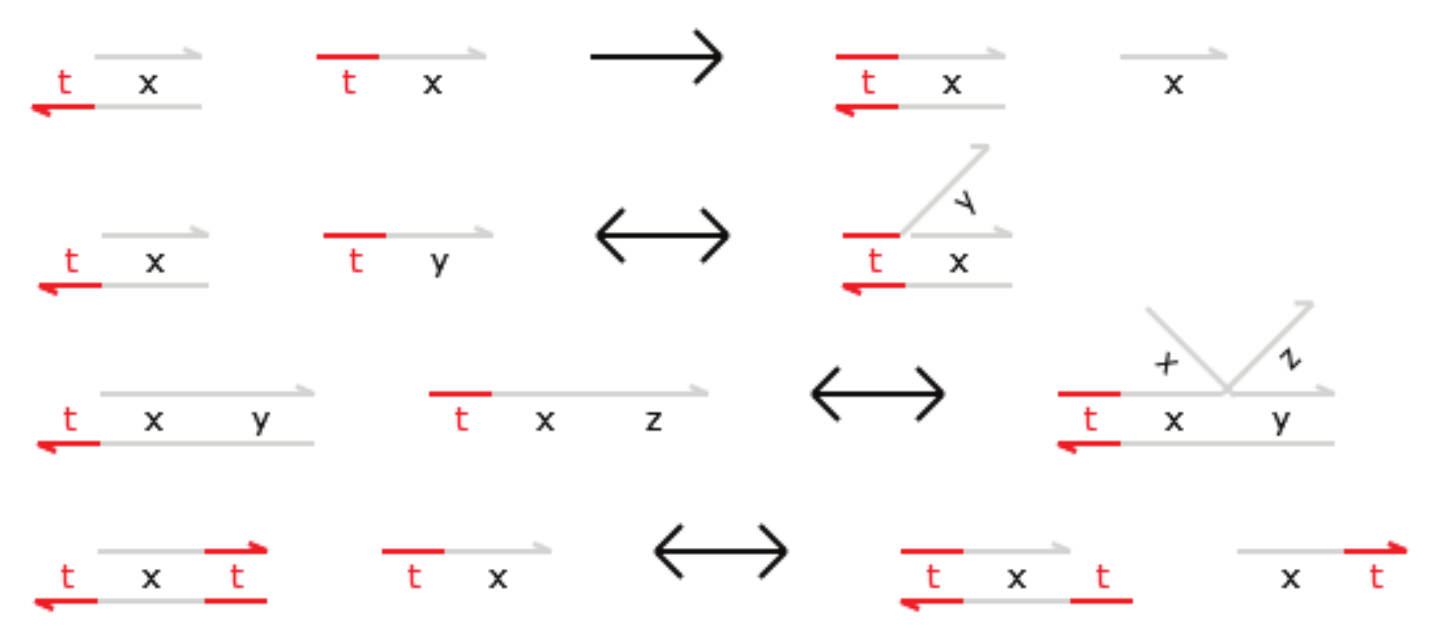}
\par\end{centering}

\caption{\label{fig:Toehold-mediated-DNA-branch}Toehold-mediated DNA branch
migration and strand displacement}

\end{figure}

The fundamental mechanism in many of these schemes is \textit{toehold
mediated branch migration and strand displacement} \cite{Yurke. Using DNA to Power Nanostructures},
which implements a basic step of computation. It operates as shown
in Figure \ref{fig:Toehold-mediated-DNA-branch}, where each letter
and corresponding segment represents a DNA \textit{domain} (a sequence
of nucleotides, $C$,$G$,$T$,$A$) and each DNA strand is seen as
the concatenation of multiple domains. Single strands have an orientation;
double strands are composed of two single strands with opposite orientation,
where the bottom strand is the Watson-Crick, $C-G$, $T-A$, complement
of the top strand. The \textquoteleft{}short\textquoteright{} domains
hybridize (bind) \textit{reversibly} to their complements, while the
\textquoteleft{}long\textquoteright{} domains hybridize \textit{irreversibly};
the exact critical length depends on physical condition. Distinct
letters indicate domains that do not hybridize with each other. 

In the first reaction of Figure \ref{fig:Toehold-mediated-DNA-branch},
a short \textit{toehold} domain $t$ initiates binding between a double
strand and a single strand. After the (reversible) binding of the
toehold, the $x$ domain of the single strand gradually replaces the
top $x$ strand of the double strand by \textit{branch migration}.
The branching point between the two top $x$ domains performs a random
walk that eventually leads to \textit{displacing} the $x$ strand.
The final detachment of the top $x$ strand makes the whole process
essentially irreversible, because there is no toehold for the reverse
reaction. The second reaction illustrates the case where the top domains
do not match: then the toehold binds reversibly and no displacement
occurs. The third reaction illustrates the more detailed situation
where the top domains matches only initially: the branch migration
can proceed only up to a certain point and then must revert back to
the toehold: hence no displacement occurs and the whole reaction reverts.

The fourth reaction illustrates a \textit{toehold exchange}, where
a branch migration (of strand $tx$) leads to a displacement (of strand
$xt$), but where the whole process is reversible via a reverse toehold
binding and branch migration. The first (irreversible) and fourth
(reversible) reactions are the fundamental steps that can be composed
to achieve computation by strand displacement.

\section{Two-domain Signals and Gates }

We now describe some DNA strand displacement structures that emulate,
depending on the point of view, either chemical reactions or Petri
net transitions. Their function is to \textit{join} input signals
and \textit{fork} output signals. To achieve compositionality, so
that gates can be composed arbitrarily into larger circuits, it is
necessary to first fix the structure of the signals. Any given choice
of signal structure requires a different gate architecture, for example
for 4-domain signals \cite{Soloveichik. DNA as a Universal Substrate for Chemical Kinetics}
(signals composed of 4 segments of different function), and 3-domain
signals \cite{Cardelli. Strand Algebras for DNA Computing}. Here
we present a new, streamlined, architecture based on 2-domain signals,
where the gates can be combined into arbitrary circuits (including
loops), and where the waste products do not interfere with the active
gates.%
\begin{figure}
\begin{centering}
\includegraphics[scale=0.6]{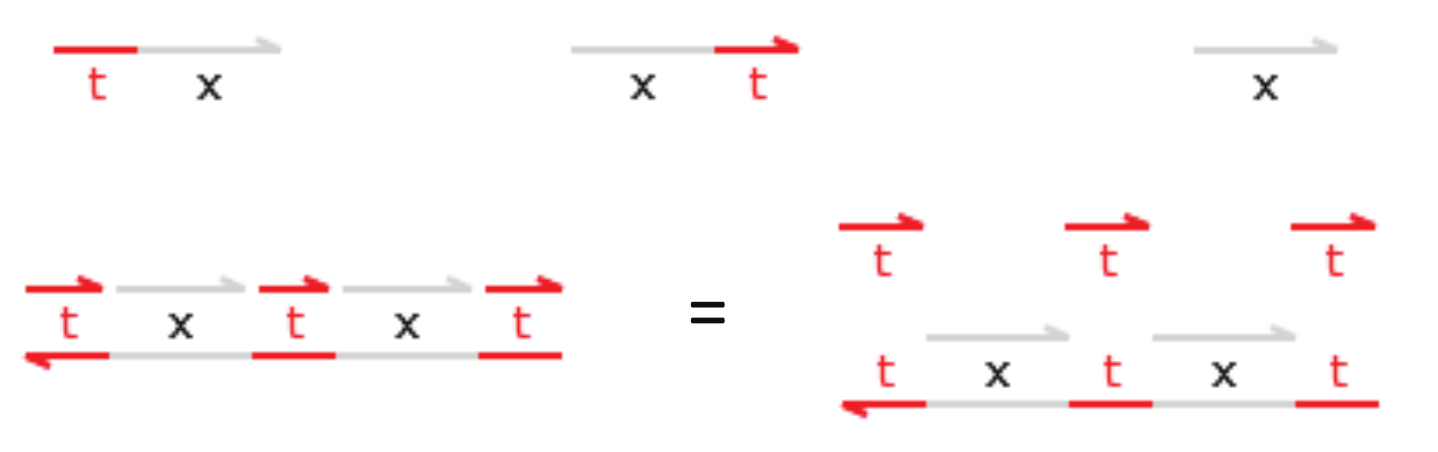}
\par\end{centering}

\caption{\label{fig:Examples-of-allowable-strands}Examples of allowable single
and double strands: $\underline{t^{\lyxmathsym{\dag}}x^{\lyxmathsym{\dag}}t^{\lyxmathsym{\dag}}x^{\lyxmathsym{\dag}}t},tx,xt,x$}

\end{figure}

\subsubsection*{Top-nicked double strands. }

Double-stranded DNA (\textit{dsDNA}) can have interruptions (\textit{nicks})
on one strand while remaining connected if the opposite strand has
enough hold on the area around the nick. We called such structures
\textit{nicked double-stranded DNA} (\textit{ndsDNA}). This excludes
any long overhangs or any protrusions from the double-strand. In particular,
we work with \textit{top-nicked double-strand}s, where all the nicks
are on one strand (the top one by convention). A deviation from this
simple structure happens fleetingly during branch migration, but all
the initial and final species we use are ndsDNA.

We use $t$ for short domains, $x$,$y$,$z$ for long domains, and
$a$,$b$,$c$ for long domains that are meant to be privately used
by some construction. We write, e.g., $tx$ for a single-stranded
DNA (\textit{ssDN}A) strand consisting of a toehold $t$ followed
by a domain $x$, and similarly for $xt$. We write, e.g., $\underline{txy}$
for a fully complemented double strand consisting of a continuous
strand $txy$ at the top and its Watson-Crick complement at the bottom.
Finally, we write $\underline{tx^{\lyxmathsym{\dag}}y}$ to indicate
the same double strand but with a nick at the top between $x$ and
$y$. In the figures, a nick is indicated by an arrowhead and a discontinuity.

Examples of allowable single and double strands are shown in Figure
\ref{fig:Examples-of-allowable-strands}. We assume that domains indicated
by different letters are distinct, so that, e.g., $x$ does not hybridize
with $y$, $zy$, $yz$, $ty$, or $yt$. To simplify our notation,
we use an implicit equivalence illustrated in the bottom part of the
figure. Suppose we start with a regular double strand, and we nick
it at the top (bottom left). Long segments between nicks remain attached
to the bottom strand, while short toehold segments can detach and
reattach (bottom right). We regard these reversible states as equivalent;
the notation $\underline{x^{\lyxmathsym{\dag}}t^{\lyxmathsym{\dag}}y}$
then indicates two equivalent situations, where the top $t$ is either
present or absent, and where $t$ is implicitly exchanged with the
environment. Hence, we can use $\underline{x^{\lyxmathsym{\dag}}t^{\lyxmathsym{\dag}}y}$
to indicate an open toehold between $x$ and $y$, because the toehold
is available (sometime). This way, we do not need to use separate
notations for temporarily occluded and temporarily open toeholds,
which we would have to regard as equivalent anyway (up to some kinetic
occlusion effects).

\subsubsection*{Two-domain strand displacement gates. }

\begin{figure}
\begin{centering}
\includegraphics[scale=0.75]{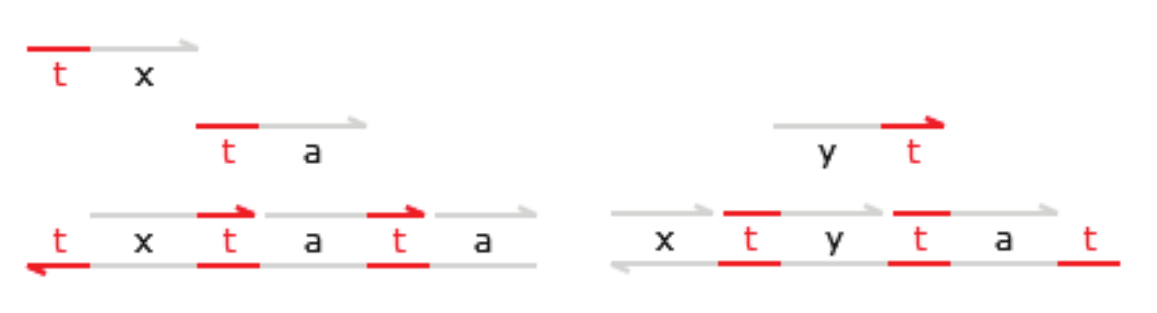}
\par\end{centering}

\caption{\label{fig:Transducer}Transducer $T_{xy}\mbox{ }|\mbox{ }tx\rightarrow ty$:
initial state plus input $tx$.}

\end{figure}

\begin{figure}
\begin{centering}
\includegraphics[scale=0.6]{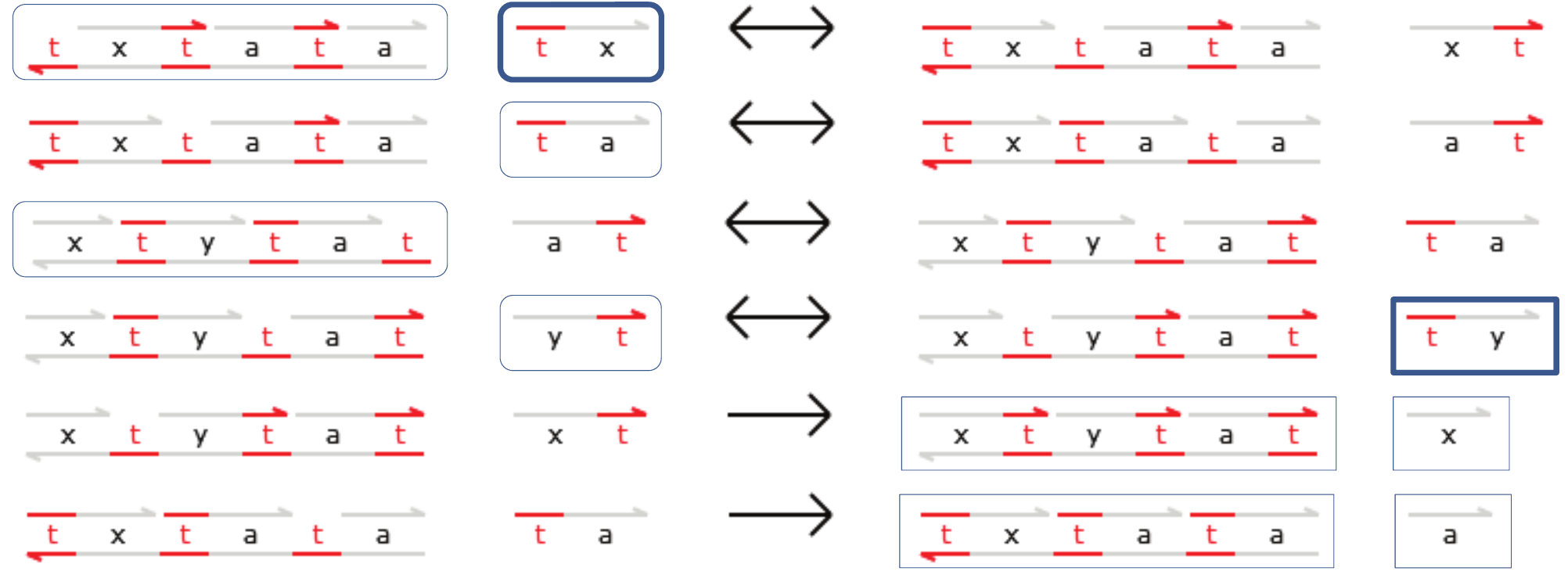}
\par\end{centering}

\caption{\label{fig:Transducer reactions}Transducer $T_{xy}\mbox{ }|\mbox{ }tx\rightarrow ty$
reactions.}

\end{figure}

All our gates are top-nicked dsDNA and our signals are two-domain
ssDNA. This simple setup is more expressive than it might appear at
first. For example (Figure \ref{fig:Transducer}), let us consider
a single strands $tx$ as encoding a \textit{signal}, with the strand
$xt$ as its \textit{cosignal}, and consider the problem of constructing
a \textit{signal transducer} $T_{xy}$ from a signal $tx$ to a signal
$ty$, with the \textit{reduction} $T_{xy}\mbox{ }|\mbox{ }tx\rightarrow ty$,
where $\mbox{ }|\mbox{ }$ is \textit{parallel composition} of components,
and final waste is discarded. All signals share the same toehold $t$,
and are distinguished by the long domains $x$,$y$,$z$, etc.%
As shown in Figure \ref{fig:Transducer reactions}, the input $tx$
can initiate a signal/cosignal cascade of strand displacements in
the left double-strand that after two toehold exchanges releases a
\textit{private} cosignal $at$ (the segment $a$ is privately used
by the $T_{xy}$ transducer, with a distinct $a$ for each $xy$ pair).
The $at$ cosignal then initiates a backward cascade in the right
double strand that releases the desired output signal $ty$ at the
fourth reaction. The release of $ty$ is reversible, but the gate
is then locked down by the last two reactions. The locking down of
the gate is also used to reabsorb the $xt$ and $ta$ strands, by
exploiting the $\underline{x}$ end of the right structure and the
$\underline{a}$ end of the left structure. In the end, only \textit{unreactive}
(no exposed toeholds) dsDNA and ssDNA is left. In Figure \ref{fig:Transducer reactions},
the initial structures from Figure \ref{fig:Transducer} are shown
inside rounded rectangles, and the final structures inside squared
rectangles. The reaction rules are described abstractly in Figure
\ref{fig:The-basic-reactions}.%

The structures in Figure \ref{fig:Transducer} can be written in the
notation described above as $T_{xy}$ $=$ $\underline{t^{\lyxmathsym{\dag}}xt^{\lyxmathsym{\dag}}at^{\lyxmathsym{\dag}}a}$
$\mbox{ }|\mbox{ }$ $ta$ $\mbox{ }|\mbox{ }$ $\underline{x^{\lyxmathsym{\dag}}ty^{\lyxmathsym{\dag}}ta^{\lyxmathsym{\dag}}t}$
$\mbox{ }|\mbox{ }$ $yt$. The auxiliary signal $ta$ contains the
private segment $a$, uniquely joining the two halves of $T_{xy}$
transducers, and we can therefore assume that it will not interfere
with other gates. The auxiliary cosignal $yt$ however contains a
public segment $y$, which is necessary to release the output signal.
It is therefore important to maintain an invariant that no other gate
in the whole system spontaneously absorbs $yt$, or in general any
public cosignal, although it may do so in a proper response to inputs.
For example, a $T_{zy}$ transducer and a $T_{xy}$ transducer may
use \textquotedblleft{}each other\textquoteright{}s\textquotedblright{}
$yt$ cosignal without problem.%
\begin{figure}
\begin{centering}
\includegraphics[scale=0.75]{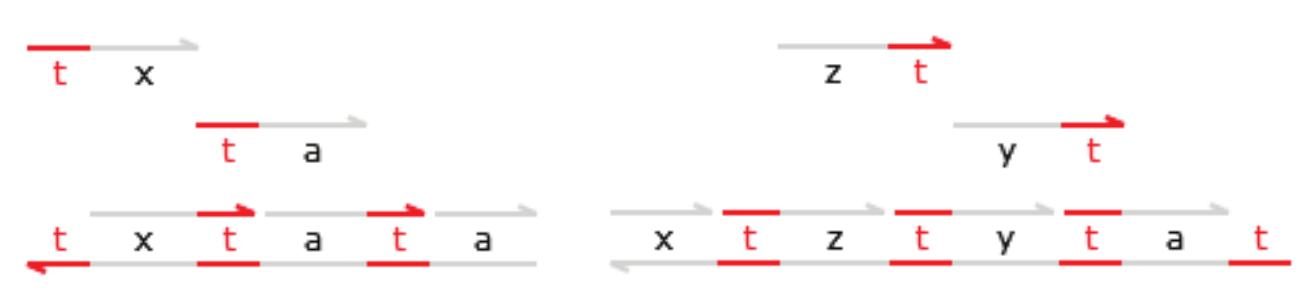}
\par\end{centering}

\caption{\label{fig:Fork}Fork $F_{xyz}\mbox{ }|\mbox{ }tx\rightarrow ty\mbox{ }|\mbox{ }tz$:
initial state plus input $tx$.}

\end{figure}

The transducer $T_{xy}$ can be extended easily to a fork gate $F_{xyz}$
such that $F_{xyz}\mbox{ }|\mbox{ }tx\rightarrow ty\mbox{ }|\mbox{ }tz$,
releasing two outputs from one input. This is shown in Figure \ref{fig:Fork},
where the left half of the structure is the same as in $T_{xy}$.%
\begin{figure}
\begin{centering}
\includegraphics[scale=0.75]{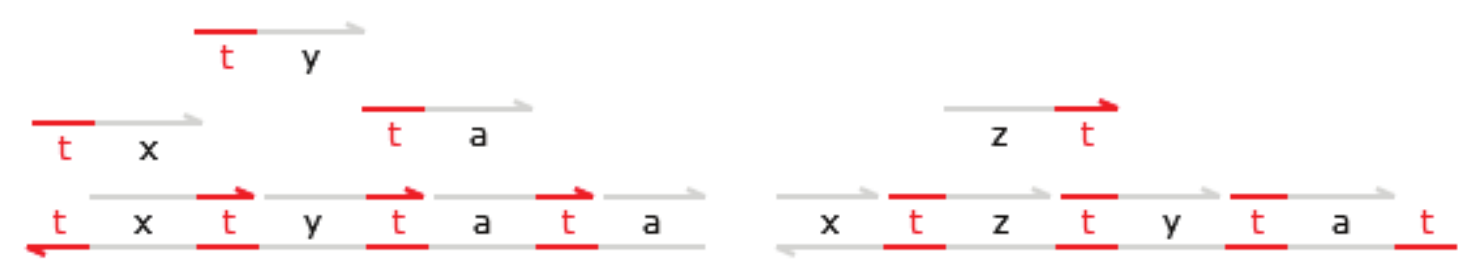}
\par\end{centering}

\caption{\label{fig:Catalyst}Catalyst $C_{xyz}\mbox{ }|\mbox{ }tx\mbox{ }|\mbox{ }ty\rightarrow ty\mbox{ }|\mbox{ }tz$:
initial state plus inputs $tx$ and $ty$.}

\end{figure}
The fork gate can be extended to a catalytic gate $C_{xyz}$ such
that $C_{xyz}\mbox{ }|\mbox{ }tx\mbox{ }|\mbox{ }ty$ $\rightarrow$
 $ty\mbox{ }|\mbox{ }tz$ (Figure \ref{fig:Catalyst}). The right
half of $C_{xyz}$ is unchanged from $F_{xyz}$, except that $yt$
is not required because it is produced by the left half. This gate,
like the more general join gate discussed next, takes two inputs,
but absorbs them only if both inputs are present \cite{Soloveichik. DNA as a Universal Substrate for Chemical Kinetics}.
If only the first input is present, it is returned to the soup by
reversibility of strand displacement between $tx$ and $xt$.%
\begin{figure}
\begin{centering}
\includegraphics[scale=0.75]{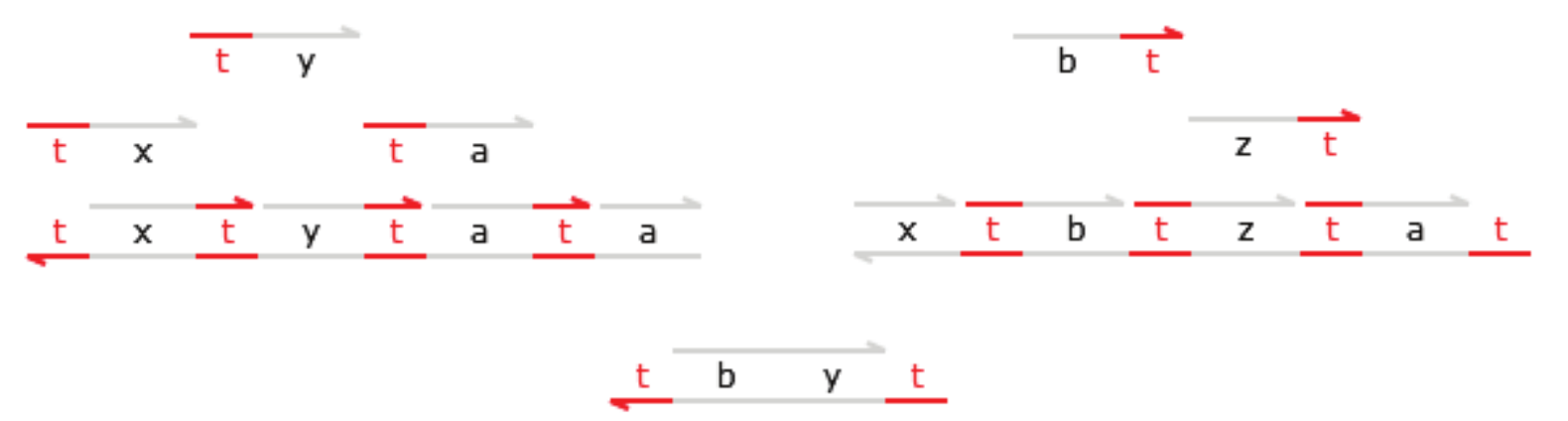}
\par\end{centering}

\caption{\label{fig:Join}Join $J_{xyz}\mbox{ }|\mbox{ }tx\mbox{ }|\mbox{ }ty\rightarrow tz$:
initial state plus inputs $tx$, $ty$.}

\end{figure}

\begin{figure}
\begin{centering}
\includegraphics[scale=0.75]{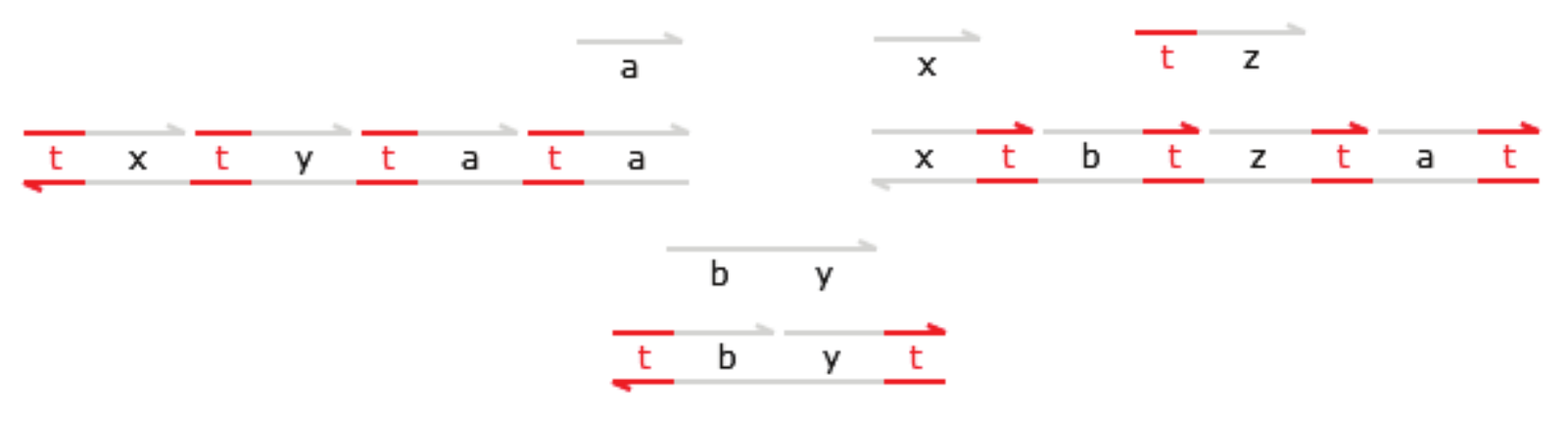}
\par\end{centering}

\caption{\label{fig:Join final state}Join $J_{xyz}\mbox{ }|\mbox{ }tx\mbox{ }|\mbox{ }ty\rightarrow tz$:
final state plus output $tz$.}

\end{figure}

\begin{figure}
\begin{centering}
\includegraphics[scale=0.6]{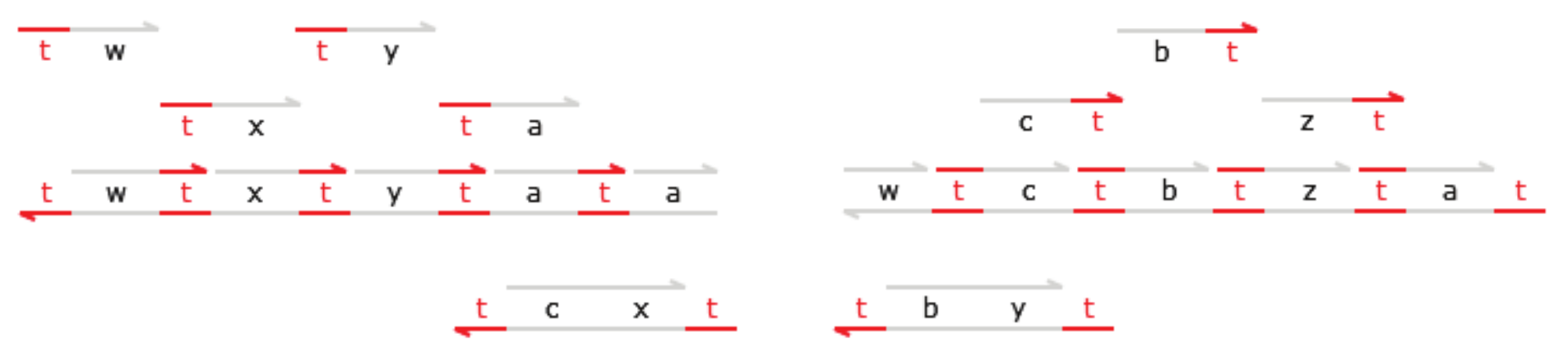}
\par\end{centering}

\caption{\label{fig:3-Join}3-Join $J_{wxyz}\mbox{ }|\mbox{ }tw\mbox{ }|\mbox{ }tx\mbox{ }|\mbox{ }ty\rightarrow tz$:
initial state plus inputs $tw$, $tx$, $ty$.}

\end{figure}

Let us now consider, in Figures \ref{fig:Join} and \ref{fig:Join final state},
a binary join gate $J_{xyz}$ such that $J_{xyz}\mbox{ }|\mbox{ }tx\mbox{ }|\mbox{ }ty$
$\rightarrow$  $tz$ (the generalization to additional outputs works
as in the fork gate). Each distinct combination of $xyz$ requires
choosing a distinct private domain connecting the two halves of the
gate; this private domain can however be shared among a population
of gates with the same input and output signals.%
The main new feature in this gate is the additional $\underline{t^{\lyxmathsym{\dag}}by^{\lyxmathsym{\dag}}t}$
structure that absorbs a signal and a cosignal together, or neither
separately. Without it, and without the $bt$, $\underline{tb}$ components,
the join gate would leave behind a $yt$ residual (all the other single
strands, $xt$, $zt$, $ta$, are reclaimed). Hence $\underline{t^{\lyxmathsym{\dag}}by^{\lyxmathsym{\dag}}t}$
is a \textquoteleft{}garbage collector\textquoteright{} turning undesired
active residuals to waste. It is triggered only after the release
of a private strand $tb$, so that the collector does not reclaim
an extraneous cosignal $yt$ before the join gate has committed to
its inputs. Such an extraneous $y$t could come from a transducer
$T_{xy}$, or from another join $J_{uvy}$ (before any input) or $J_{yuv}$
(after the first input) causing cross-gate interference, or even from
within the same join, as in $J_{xyy}$. Removing garbage is important
because accumulated garbage slows down future reactions by imposing
a growing reverse pressure on the desired direction of the reactions.
We have designed all gates to remove all active garbage, but until
now garbage removal did not require additional double strands. The
Join structure is easily generalized to any number of inputs; for
example, Figure \ref{fig:3-Join} shows a 3-input Join with collectors.%

\subsubsection*{Discussion: The double strand restrictions. }

The restriction of allowing only ndsDNA structures has a number of
potential advantages. The absence of any branching seems inherently
more trouble-free than complex structures that can interact in unexpected
ways through their protruding single-stranded parts. Here all double-stranded
structures are quiescent (except for receptive toeholds on the bottom
strand) and only single-stranded components have hybridization potential,
eliminating the possibility that the gate themselves may polymerize,
or may self-interact. These structures also have a simple syntactical
representation and simple reduction rules, which simplify formal verification.
Nothing prevents us from devising precise syntax and reductions for
more general structures \cite{Phillips. A Programming Language for Composable DNA Circuits},
and there is no good reason in principle to avoid more complex structures
if they work well. However, we have shown that our simplified structures
already cover a surprising range of computation (fork and join gates
in populations are equivalent to Petri Nets \cite{Cardelli. Strand Algebras for DNA Computing}),
and hence one can restrict the use of more complex structures to the
situations where they are actually needed, or where they somehow perform
better.

\subsubsection*{Discussion: The single strand restrictions. }

Our hybridized structures start as ndsDNA, but we have to ensure that
they remain ndsDNA through computation. (Except for \textit{transients,}
i.e., during branch migrations that either revert harmlessly or lead
to strand displacements.) This invariant puts constraints on the allowable
single strands. First of all, single strands consisting only of long
segments are inert because all the double strands are fully complemented
(except for toeholds), and hence they can be ignored. A single strand
of the form $xty$ could bind to a double strand of the form $\underline{x^{\lyxmathsym{\dag}}t^{\lyxmathsym{\dag}}z}$,
leading to a configuration that is stable and is not ndsDNA. Therefore
our single strands cannot contain substrands of the form $xty$, and
we are left with single strands of the form, $x^{n}t^{m}$ or $t^{n}x^{m}$
or $t^{n}x^{m}t^{p}$. The third class could lead to stable configurations
with two overlapping competing toeholds ($\underline{t^{\lyxmathsym{\dag}}x^{\lyxmathsym{\dag}}t^{\lyxmathsym{\dag}}y^{\lyxmathsym{\dag}}t}$
with $txt$ and $tyt$) and hence are ruled out too. Multiple toeholds
in sequence bind as stably as a long domain, so e.g. $xttt$ would
be as bad as the former $xty$, and they can lead to competing toeholds:
$\underline{x^{\lyxmathsym{\dag}}t^{\lyxmathsym{\dag}}t^{\lyxmathsym{\dag}}y}$
with $xtt$ and $tty$. Hence we do not allow consecutive toeholds
in the top strands. Similarly, strands with consecutive long segments
can lead to stable competition: $txy$ and $yzt$ over $\underline{t^{\lyxmathsym{\dag}}xyz^{\lyxmathsym{\dag}}t}$.
In the end, we are left only with $xt$ or $tx$, and the only remaining
competition is between $tx$ and $xt$ over $\underline{t^{\lyxmathsym{\dag}}x^{\lyxmathsym{\dag}}t}$,
where the stable structures are ndsDNA. A final case to consider is
$tx$ and $yt$ over $\underline{t^{\lyxmathsym{\dag}}xy^{\lyxmathsym{\dag}}t}$:
if a single strand is present it binds only reversibly, and if both
are present they both bind stably and release $xy$, so the stable
structures are always ndsDNA. In fact, $\underline{t^{\lyxmathsym{\dag}}xy^{\lyxmathsym{\dag}}t}$
is an important configuration that seems to add some power: without
it we can still implement garbage-collecting join gates, but apparently
only by using more than one distinct toehold.

\subsubsection*{Discussion: The double strand restrictions, revisited. }

We finally have to make sure that no reactive single strands other
than $t$, $tx$, $xt$, plus the unreactive $x$ and $xy$, are ever
released from double strands during computation. This imposes another
restriction on double strands: nicks should break the top strand into
segments of two domains or less. Otherwise, the double strand $\underline{t^{\lyxmathsym{\dag}}xty^{\lyxmathsym{\dag}}t}$
could release a forbidden single strand $xty$ in presence of $tx$
and $yt$. (We could still allow $\underline{t^{\lyxmathsym{\dag}}xyz^{\lyxmathsym{\dag}}t}$,
but it would be unreactive.) Hence, we are left with allowable double
strands that are nicked concatenations of the double-stranded elements
$\underline{t}$, $\underline{x}$, $\underline{tx}$, $\underline{xt}$,
$\underline{xy}$.

\section{Nick Algebra }

In this section we provided a formal framework where we can perform
calculations about the evolution of systems of top-nicked double strands.
\textit{Domains} are taken either from a finite set of \textit{short
domains} (\textit{toeholds}) or from an unbounded set of \textit{long
domains} ranged over by $x$,$y$,$z$ and $a$,$b$,$c$. The set
of toeholds must be finite (and in practice quite small) because of
its reversible-binding assumption that limits length and hence cardinality.
Designs based on a single toehold can be easily adapted to multiple
toeholds to increase binding discrimination and efficiency, but the
converse is problematic: designs based on distinct toeholds may fail
if the toeholds are then identified. Here we require only a single
distinguished toehold, always indicated by the constant $t$, but
it would be easy to generalize to multiple toeholds. 

An infix operator \textquoteleft{}$.$\textquoteright{} may be used
to concatenate domains into single strands; this is often omitted,
particularly because all our single-strands have the form $t.x$ or
$x.t$, which are then usually written $tx$ and $xt$ (unless we
wish to use long identifiers for domains). Single strands $t$, $x$,
and $x.y$ remain implicit \textquoteleft{}waste\textquoteright{},
and are not used in the syntax. %
\begin{figure}
\begin{centering}
\includegraphics[scale=0.58]{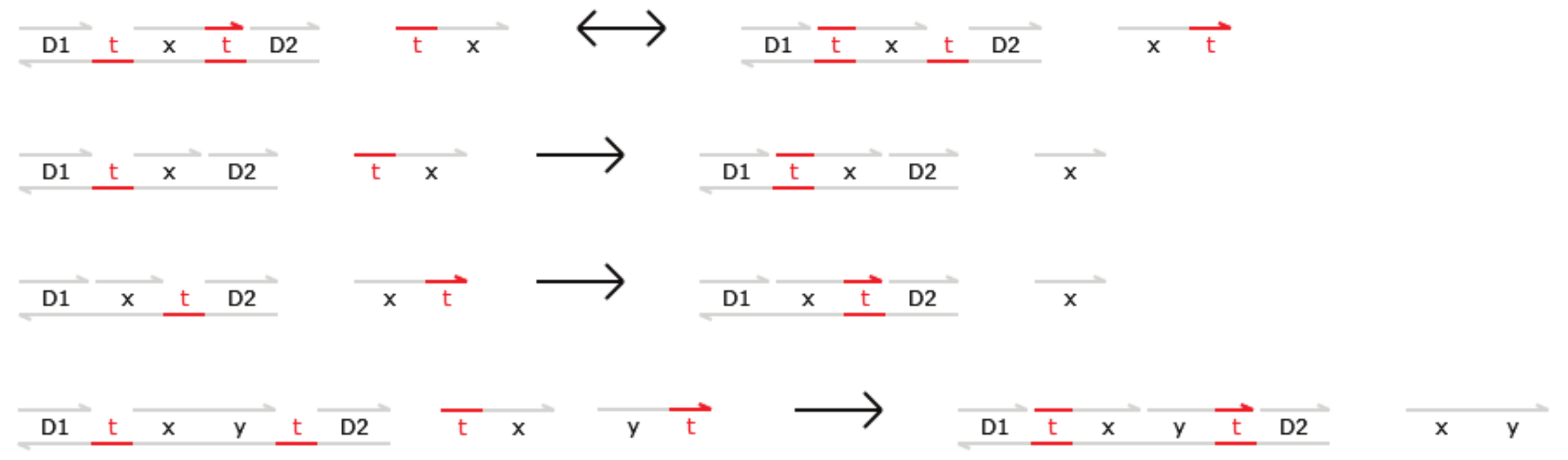}
\par\end{centering}

\caption{\label{fig:The-basic-reactions}The basic reactions (D1,D2 are arbitrary
or empty double strands).}

\end{figure}

Double strands are written underlined. We use an infix operator \textquoteleft{}$\underline{^{\lyxmathsym{\dag}}}$\textquoteright{}
to represent a \textquoteleft{}nick\textquoteright{} on the top strand
of a double-stranded sequence, an infix operator \textquoteleft{}$\underline{.}$\textquoteright{}
(often omitted) to represent the unbroken concatenation of top and
bottom strands, and $\slashed{o}$ for the empty double strand. The
segments between nicks are only single or pair combinations of toeholds
and domains. 

A soup $U$ is a finite multiset of single and double strands, with
multiset union indicated by \textquoteleft{}$\mbox{ }|\mbox{ }$\textquoteright{},
and with a notation $(\nu x)U$ for domain isolation. The latter indicates
that $x$ is not used outside of $U$: this allows us to declare private
domains locally, and to combine constructions compositionally. In
practice, it means simply that all the domains indicated by $\nu$
must be chosen distinct when a global system is fixed for execution:
the algebraic laws for $(\nu x)U$ encode such a guarantee. We also
use $U^{n}$ as an abbreviation for $n$ copies of $U$ in parallel
($\mbox{ }|\mbox{ }$). The resulting algebra is our nick algebra,
which is strictly a subset of the DSD (DNA Strand Displacement) language
\cite{Phillips. A Programming Language for Composable DNA Circuits}.

\subsubsection*{Definition: Term Syntax}

\vspace{0in}

\begin{tabular}{>{\raggedright}p{3in}>{\raggedright}p{2in}}
$S::=t.x\mbox{ }\brokenvert\mbox{ }x.t$

$\underline{D}::=\slashed{o}\mbox{ }\brokenvert\mbox{ }\underline{t}\mbox{ }\brokenvert\mbox{ }\underline{x}\mbox{ }\brokenvert\mbox{ }\underline{t.x}\mbox{ }\brokenvert\mbox{ }\underline{x.t}\mbox{ }\brokenvert\mbox{ }\underline{x.x}\mbox{ }\brokenvert\mbox{ }\underline{D^{\text{\dag}}D}$

$U::=S\mbox{ }\brokenvert\mbox{ }\underline{D}\mbox{ }\brokenvert\mbox{ }U|U\mbox{ }\brokenvert\mbox{ }(\nu x)U$ & Single strand

Double strand

Soup\tabularnewline
\end{tabular}

\medskip{}

The set of \textit{public domains} $pd(U)$ is the inductively defined
set of those domains not bound by $\nu$ in $U$; in particular $pd(t.x)$
= $pd(x.t)$ $=$ $pd(\underline{x})$ $=$ $pd(\underline{t.x})$
$=$ $pd(\underline{x.t})$ $=$ $\{x\}$, $pd(\underline{x.y})$
$=$ $\{x,y\}$, $pd(\underline{t})$ $=$ $pd(\slashed{o})$ $=$
$\{\}$, and $pd((\nu x)U)$ $=$ $pd(U)-\{x\}$. Then, $U\{y/x\}$
is the substitution of $y$ for $x$ in $U$, with the representative
cases $t\{y/x\}=t$, $x\{y/x\}=y$, $z\{y/x\}=z$ for $z\neq x$,
$((\nu z)U)\{y/x\}$ $=$ $(\nu z)U\{y/x\}$ for $z\notin\{x,y\}$,
$((\nu x)U)\{y/x\}$ $=$ $(\nu x)U$, and $((\nu y)U)\{y/x\}$ $=$
$(\nu z)U\{z/y\}\{y/x\}$ for a $z\notin pd(U)\cup\{x,y\}$. 

\textit{Algebraic equality} (a binary congruence relation over the
term syntax) is indicated just by $=$ and is axiomatized below with
the monoid laws of $(\slashed{o},\underline{^{\lyxmathsym{\dag}}})$,
the commutative monoid laws of $(\slashed{o},\mbox{ }|\mbox{ })$,
and the scoping laws of $(\nu x)U$ \cite{Milner. The pi-Calculus}.

\subsubsection*{Definition: Algebraic Equality}

\vspace{0in}

$=$ is an equivalence relation

\medskip{}

\begin{minipage}[t]{1\columnwidth}%
$\underline{D_{1}}=\underline{D_{2}},\mbox{ }\mbox{ }\underline{D_{3}}=\underline{D_{4}}\mbox{ }\mbox{ }\Rightarrow\mbox{ }\mbox{ }\underline{D_{1}{}^{\text{\dag}}D_{3}}=\underline{D_{2}{}^{\text{\dag}}D_{4}}$

$U_{1}=U_{2},\mbox{ }\mbox{ }U_{3}=U_{4}\mbox{ }\mbox{ }\Rightarrow\mbox{ }\mbox{ }U_{1}\mbox{ }|\mbox{ }U_{3}\mbox{ }=\mbox{ }U_{2}\mbox{ }|\mbox{ }U_{4}$

$U_{1}=U_{2}\mbox{ }\mbox{ }\Rightarrow\mbox{ }\mbox{ }(\text{\ensuremath{\nu}}x)U_{1}\mbox{ }=\mbox{ }(\text{\ensuremath{\nu}}x)U_{2}$%
\end{minipage}

\medskip{}

\begin{minipage}[t]{1\columnwidth}%
$\underline{D_{1}{}^{\lyxmathsym{\dag}}(D_{2}{}^{\lyxmathsym{\dag}}D_{3})}=\underline{(D_{1}{}^{\lyxmathsym{\dag}}D_{2})^{\lyxmathsym{\dag}}D_{3}}$

$\slashed{o}\underline{{}^{\lyxmathsym{\dag}}D}=\underline{D{}^{\lyxmathsym{\dag}}}\slashed{o}=\underline{D}$%
\end{minipage}

\medskip{}

\begin{minipage}[t]{1\columnwidth}%
$U_{1}\mbox{ }|\mbox{ }(U_{2}\mbox{ }|\mbox{ }U_{3})=(U_{1}\mbox{ }|\mbox{ }U_{2})\mbox{ }|\mbox{ }U_{3}$

$U_{1}\mbox{ }|\mbox{ }U_{2}=U_{2}\mbox{ }|\mbox{ }U_{1}$

$\slashed{o}\mbox{ }|\mbox{ }U=U\mbox{ }|\mbox{ }\slashed{o}=U$%
\end{minipage}

\medskip{}

\begin{minipage}[t]{1\columnwidth}%
$(\nu x)U=(\nu y)(U\{y/x\})$$\qquad$if $y\notin pd(U)$

$(\nu x)\slashed{o}=\slashed{o}$

$(\nu x)(U_{1}\mbox{ }|\mbox{ }U_{2})=U_{1}\mbox{ }|\mbox{ }(\nu x)U_{2}$$\qquad$if
$x\notin pd(U_{1})$

$(\nu x)(\nu y)U=(\nu y)(\nu x)U$%
\end{minipage}

\medskip{}

Note that $(\nu x)(\nu x)U=(\nu x)U$ is derivable. As an example
of use of the isolation operation, consider that it is always possible
to bring all the $\nu$ prefixes to the top level by making all the
private domains distinct: $(\nu x)tx$ $\mbox{ }|\mbox{ }$ $(\nu x)tx=(\nu x)tx$
$\mbox{ }|\mbox{ }$ $(\nu y)ty=(\nu x)(\nu y)(tx$ $\mbox{ }|\mbox{ }$
$ty)$. This means that conflicts between local definitions can be
resolved globally, while allowing local definition to be combined
without consideration of global conflicts. 

The\textit{ reduction} relation $U_{1}\rightarrow U_{2}$ describes
a single step of system evolution; it is the smallest binary relation
on $U$ satisfying the rules below, where $\leftrightarrow$ stands
for two reduction rules in opposite directions. Its symmetric and
transitive closure $U_{1}\rightarrow^{*}U_{2}$ describes multi-step
system evolution. In the reduction rules, the \textit{single-stranded
waste} ($t$, $x$, $xy$) is automatically removed because it can
be immediately identified as waste (as a consequence, the single strands
$t$, $x$, $xy$ need not be included in the syntax). Alternatively,
we could have made the single-stranded waste explicit and introduced
separate rules to remove it. The double-stranded waste instead has
a special degradation rule because it requires a check over the whole
double strand. The four basic reactions (exchange, coverage, cooperation)
are depicted in Figure \ref{fig:The-basic-reactions}.

\subsubsection*{Definition: Reduction}

\vspace{0in}

\begin{tabular}{>{\raggedright}p{3in}>{\raggedright}p{2in}}
$\underline{D_{1}{}^{\lyxmathsym{\dag}}t^{\lyxmathsym{\dag}}xt^{\lyxmathsym{\dag}}D_{2}}\mbox{ }|\mbox{ }tx\leftrightarrow\underline{D_{1}{}^{\lyxmathsym{\dag}}tx^{\lyxmathsym{\dag}}t^{\lyxmathsym{\dag}}D_{2}}\mbox{ }|\mbox{ }xt$

$\underline{D_{1}{}^{\lyxmathsym{\dag}}t^{\lyxmathsym{\dag}}x^{\lyxmathsym{\dag}}D_{2}}\mbox{ }|\mbox{ }tx\rightarrow\underline{D_{1}{}^{\lyxmathsym{\dag}}tx^{\lyxmathsym{\dag}}D_{2}}$

$\underline{D_{1}{}^{\lyxmathsym{\dag}}x^{\lyxmathsym{\dag}}t^{\lyxmathsym{\dag}}D_{2}}\mbox{ }|\mbox{ }xt\rightarrow\underline{D_{1}{}^{\lyxmathsym{\dag}}xt^{\lyxmathsym{\dag}}D_{2}}$

$\underline{D_{1}{}^{\lyxmathsym{\dag}}t^{\lyxmathsym{\dag}}xy^{\lyxmathsym{\dag}}t^{\lyxmathsym{\dag}}D_{2}}\mbox{ }|\mbox{ }tx\mbox{ }|\mbox{ }yt\rightarrow\underline{D_{1}{}^{\lyxmathsym{\dag}}tx^{\lyxmathsym{\dag}}yt^{\lyxmathsym{\dag}}D_{2}}$ & Exchange

Left coverage

Right coverage

Cooperation \tabularnewline
\end{tabular}

\medskip{}

\begin{tabular}{>{\raggedright}p{3in}>{\raggedright}p{2in}}
$\underline{D}\rightarrow\slashed{o}$$\qquad$if $\underline{D}$
not \textit{reactive} 

$U_{1}\rightarrow U_{2}\mbox{ }\mbox{ }\Rightarrow\mbox{ }\mbox{ }U_{1}\mbox{ }|\mbox{ }U\rightarrow U_{2}\mbox{ }|\mbox{ }U$

$U_{1}\rightarrow U_{2}\mbox{ }\Rightarrow\mbox{ }(\nu x)U_{1}\rightarrow(\nu x)U_{2}$

$U_{1}=U_{2},U_{2}\rightarrow U_{3},U_{3}=U_{4}\mbox{ }\Rightarrow\mbox{ }U_{1}\rightarrow U_{4}$ & Waste

Dilution

Isolation 

Well-mixing\tabularnewline
\end{tabular}

\medskip{}

A double strand $\underline{D}$ is \textit{reactive} if it can react
in some context; that is, by the first four rules. Hence it must be
of the form $\underline{D_{1}{}^{\lyxmathsym{\dag}}t^{\lyxmathsym{\dag}}xt^{\lyxmathsym{\dag}}D_{2}}$,
$\underline{D_{1}{}^{\lyxmathsym{\dag}}tx^{\lyxmathsym{\dag}}t^{\lyxmathsym{\dag}}D_{2}}$,
$\underline{D_{1}{}^{\lyxmathsym{\dag}}t^{\lyxmathsym{\dag}}x^{\lyxmathsym{\dag}}D_{2}}$,
$\underline{D_{1}{}^{\lyxmathsym{\dag}}x^{\lyxmathsym{\dag}}t^{\lyxmathsym{\dag}}D_{2}}$,
or $\underline{D_{1}{}^{\lyxmathsym{\dag}}t^{\lyxmathsym{\dag}}xy^{\lyxmathsym{\dag}}t^{\lyxmathsym{\dag}}D_{2}}$.
Among the unreactive (waste) double strands are thus $\underline{t}$,
$\underline{x}$, $\underline{xt}$, $\underline{tx}$, $\underline{xy}$,
$\underline{t^{\lyxmathsym{\dag}}t}$, $\underline{t^{\lyxmathsym{\dag}}tx}$,
$\underline{xt^{\lyxmathsym{\dag}}t}$, $\underline{xt^{\lyxmathsym{\dag}}ty}$,
$\underline{xt^{\lyxmathsym{\dag}}t^{\lyxmathsym{\dag}}ty}$, etc.
The waste rule is really a convenience to simplify results of calculations;
more generally, as commonly done in process algebra, one would instead
eliminate unreactive components via an observational equivalence \cite{Milner. The pi-Calculus}.

\section{Correctness }

If $U_{1}\rightarrow^{*}U_{2}$ then $U_{1}$ may reduce to $U_{2}$,
but it may also reduce to something else since $\rightarrow^{*}$
is a relation. When $U_{1}\rightarrow^{*}U_{2}$ is used to state
a correctness property of system reduction, we say that this is a
\textit{may-correctness} property: the system starting from $U_{1}$
\textit{may} reduce to $U_{2}$, but it may also wander in a different
section of state space and never be able to get to $U_{2}$ from there.
A stronger property is \textit{will-correctness}, indicated by $U_{1}\rightarrow^{\forall}U_{2}$,
and defined as $\forall U,$ $U_{1}\rightarrow^{*}U$ $\mbox{ }\Rightarrow\mbox{ }$
$U\rightarrow^{*}U_{2}$. This means that although $U_{1}$ may wander
to some $U$ in some part of the state space, it \textit{will} always
find a path to $U_{2}$ from there (it \textit{cannot avoid} finding
a path to $U_{2}$). If $U_{1}\rightarrow^{\forall}U_{2}$ and $U_{2}$
is the only terminal state, then we can say that $U_{1}$ \textit{must
reduce} to $U_{2}$. But will-correctness does not imply that reduction
necessarily terminates, and in particular if $U\rightarrow^{\forall}U$
we can say that $U$ is \textit{reversible}. Since $U_{1}\rightarrow^{*}U_{1}$
holds by reflexivity, will-correctness implies may-correctness. (All
these properties are really examples of a large class of reachability
properties that could be expressed in a temporal logic.)

It is convenient in the next examples and proofs to use a more pictographic
notation for nick algebra expressions, to highlight the positions
of the toeholds. We use the following abbreviations ($\underline{^{\lyxmathsym{\dag}}}$
is still needed in for $\underline{x^{\lyxmathsym{\dag}}y}$):

\subsubsection*{Definition: Two-Domain Pictograms }

\vspace{0in}

\begin{tabular}{>{\raggedright}p{0.5in}>{\raggedright}p{2.8in}>{\raggedright}p{1.5in}}
$_{\ulcorner}x$

$x_{\urcorner}$

$\underline{D{}_{\ulcorner}x}$

$\underline{x_{\urcorner}D}$

$\underline{D_{\smile}D\text{\textquoteright}}$ & for $tx$

for $xt$

for $\underline{D^{\lyxmathsym{\dag}}tx}$ (including $D=\slashed{o}$)

for $\underline{xt^{\lyxmathsym{\dag}}D}$ (including $D=\slashed{o}$)

for $\underline{D^{\lyxmathsym{\dag}}t^{\lyxmathsym{\dag}}D\lyxmathsym{\textquoteright}}$
(including $D=\slashed{o}$ or $D\lyxmathsym{\textquoteright}=\slashed{o}$) & \textit{Signal}

\textit{Cosignal}

Bound signal

Bound cosignal

Bottom toehold \tabularnewline
\end{tabular}

\medskip{}

For example, the transducer from Figure \ref{fig:Transducer} can
be written as:

\medskip{}

\begin{tabular}{lll}
$\underline{t^{\lyxmathsym{\dag}}xt^{\lyxmathsym{\dag}}at^{\lyxmathsym{\dag}}a}\mbox{ }|\mbox{ }ta\mbox{ }|\mbox{ }\underline{x^{\lyxmathsym{\dag}}ty^{\lyxmathsym{\dag}}ta^{\lyxmathsym{\dag}}t}\mbox{ }|\mbox{ }yt$ & $\ \ \ \ $ & explicit notation \tabularnewline
$\underline{_{\smile}x_{\urcorner}a_{\urcorner}a}\mbox{ }|\mbox{ }{}_{\ulcorner}a\mbox{ }|\mbox{ }\underline{x{}_{\ulcorner}y{}_{\ulcorner}a_{\smile}}\mbox{ }|\mbox{ }y_{\urcorner}$  &  & pictogram notation\tabularnewline
\end{tabular}

\medskip{}

We now show that the transducer \textit{may} work correctly. Because
of their chemical origin, all components come in populations of identical
molecules, and any private domain can only be private to a population,
and not to an individual molecule. Hence we need to show that a populations
of transducers, all sharing the same private domain, \textit{may}
map an input population to a desired output population.

\subsubsection*{\label{pro:Transducer-May-Correctness}Proposition 1: Transducer
$T_{xy}^{n}$ May-Correctness }

\vspace{0in}

Let $T_{xy}^{n}$ $=$ $(\nu a)((\underline{_{\smile}x_{\urcorner}a_{\urcorner}a}\mbox{ }|\mbox{ }_{\ulcorner}a\mbox{ }|\mbox{ }\underline{x_{\ulcorner}y_{\ulcorner}a_{\smile}}\mbox{ }|\mbox{ }y_{\urcorner})^{n})$,

then $T_{xy}^{n}\mbox{ }|\mbox{ }_{\ulcorner}x^{n}\rightarrow^{*}{}_{\ulcorner}y^{n}$.

\subsubsection*{Proof}

Let $T{}_{xay}$ = $\underline{_{\smile}x_{\urcorner}a_{\urcorner}a}\mbox{ }|\mbox{ }_{\ulcorner}a\mbox{ }|\mbox{ }\underline{x_{\ulcorner}y_{\ulcorner}a_{\smile}}\mbox{ }|\mbox{ }y_{\urcorner}$
for $a\neq x,y$, so that $T_{xy}^{n}=(\text{\ensuremath{\nu}}a)((T{}_{xay})^{n})$.
We first show that $T{}_{xay}\mbox{ }|\mbox{ }_{\ulcorner}x\rightarrow^{*}{}_{\ulcorner}y$.

$T{}_{xay}\mbox{ }|\mbox{ }_{\ulcorner}x$

$=\underline{_{\smile}x_{\urcorner}a_{\urcorner}a}\mbox{ }|\mbox{ }_{\ulcorner}a\mbox{ }|\mbox{ }\underline{x_{\ulcorner}y_{\ulcorner}a_{\smile}}\mbox{ }|\mbox{ }y_{\urcorner}\mbox{ }|\mbox{ }_{\ulcorner}x$

$\leftrightarrow\underline{_{\ulcorner}x_{\smile}a_{\urcorner}a}\mbox{ }|\mbox{ }_{\ulcorner}a\mbox{ }|\mbox{ }\underline{x_{\ulcorner}y_{\ulcorner}a_{\smile}}\mbox{ }|\mbox{ }y_{\urcorner}\mbox{ }|\mbox{ }x_{\urcorner}$

$\leftrightarrow\underline{_{\ulcorner}x_{\ulcorner}a_{\smile}a}\mbox{ }|\underline{\mbox{ }x_{\ulcorner}y_{\ulcorner}a_{\smile}}\mbox{ }|\mbox{ }y_{\urcorner}\mbox{ }|\mbox{ }x_{\urcorner}\mbox{ }|\mbox{ }a_{\urcorner}$

$\leftrightarrow\underline{_{\ulcorner}x_{\ulcorner}a_{\smile}a}\mbox{ }|\mbox{ }\underline{x_{\ulcorner}y_{\smile}a_{\urcorner}}\mbox{ }|\mbox{ }y_{\urcorner}\mbox{ }|\mbox{ }x_{\urcorner}\mbox{ }|\mbox{ }_{\ulcorner}a$

$\rightarrow\underline{_{\ulcorner}x_{\ulcorner}a_{\ulcorner}a}\mbox{ }|\underline{\mbox{ }x_{\ulcorner}y_{\smile}a_{\urcorner}}\mbox{ }|\mbox{ }y_{\urcorner}\mbox{ }|\mbox{ }x_{\urcorner}$

$\rightarrow\underline{x_{\ulcorner}y_{\smile}a_{\urcorner}}\mbox{ }|\mbox{ }y_{\urcorner}\mbox{ }|\mbox{ }x_{\urcorner}$

$\leftrightarrow\underline{x_{\smile}y_{\urcorner}a_{\urcorner}}\mbox{ }|\mbox{ }x_{\urcorner}\mbox{ }|\mbox{ }_{\ulcorner}y$

$\rightarrow\underline{x_{\urcorner}y_{\urcorner}a_{\urcorner}}\mbox{ }|\mbox{ }_{\ulcorner}y$

$\rightarrow{}_{\ulcorner}y$

Hence $(T{}_{xay}\mbox{ }|\mbox{ }_{\ulcorner}x)^{n}$ $\rightarrow^{*}$
$_{\ulcorner}y^{n}$ by induction, $(T{}_{xay})^{n}$ $\mbox{ }|\mbox{ }$
$_{\ulcorner}x^{n}$ $\rightarrow^{*}$ $_{\ulcorner}y^{n}$ by associativity,
$(\nu a)((T{}_{xay})^{n}$ $\mbox{ }|\mbox{ }$ $_{\ulcorner}x^{n})$
$\rightarrow^{*}$ $(\nu a)_{\ulcorner}y^{n}$ by isolation, and $T_{xy}^{n}\mbox{ }|\mbox{ }_{\ulcorner}x^{n}$
$\rightarrow^{*}$ $_{\ulcorner}y^{n}$ by $\nu$-equivalence and
by $T_{xy}^{n}$ definition. \textbf{End proof.}

We can similarly check the may-correctness of fork and join gates:

\subsubsection*{\label{pro:Fork-May-Correctness}Proposition 2: Fork $F_{xyz}^{n}$
May-Correctness }

\vspace{0in}

Let $F_{xyz}^{n}=(\nu a)((\underline{_{\smile}x_{\urcorner}a_{\urcorner}a}\mbox{ }|\mbox{ }_{\ulcorner}a\mbox{ }|\mbox{ }\underline{x_{\ulcorner}z_{\ulcorner}y_{\ulcorner}a_{\smile}}\mbox{ }|\mbox{ }z_{\urcorner}\mbox{ }|\mbox{ }y_{\urcorner})^{n})$,

then $F_{xyz}^{n}\mbox{ }|\mbox{ }_{\ulcorner}x^{n}\rightarrow^{*}{}_{\ulcorner}y^{n}\mbox{ }|\mbox{ }_{\ulcorner}z^{n}$.

\medskip{}

\subsubsection*{\textmd{\label{pro:Join-May-Correctness}}Proposition 3: Join \textmd{$J_{xyz}^{n}$}
May-Correctness }

\vspace{0in}

Let $J_{xyz}^{n}=(\nu a)(\nu b)((\underline{_{\smile}x_{\urcorner}y_{\urcorner}a_{\urcorner}a}\mbox{ }|\mbox{ }_{\ulcorner}a\mbox{ }|\mbox{ }\underline{x_{\ulcorner}b_{\ulcorner}z_{\ulcorner}a_{\smile}}\mbox{ }|\mbox{ }b_{\urcorner}\mbox{ }|\mbox{ }z_{\urcorner}\mbox{ }|\mbox{ }\underline{_{\smile}b^{\lyxmathsym{\dag}}y_{\smile}})^{n})$,

then $J_{xyz}^{n}\mbox{ }|\mbox{ }_{\ulcorner}x^{n}\mbox{ }|\mbox{ }_{\ulcorner}y^{n}\rightarrow^{*}{}_{\ulcorner}z^{n}$.

\medskip{}

Consider now the difficulties involved in proving more interesting
properties. We would like a transducer, for example, to work correctly
in \textquoteleft{}all possible contexts\textquoteright{}. Unfortunately
that is just not true, because some context could absorb the $y_{\urcorner}$
strand, which is public, and interfere with the transducer. One would
have to consider instead \textquoteleft{}all possible contexts that
do not interfere with $y_{\urcorner}$\textquoteright{}. This is a
rather awkward notion: for compositionality one would have, for each
component, to keep track of all the elements in the context that the
component might be interfering with. Moreover, the transducer interferes
with $y_{\urcorner}$, and hence it interferes with (another copy
or another population of) itself.

Let us consider a simpler \textquoteleft{}progress\textquoteright{}
property: that the transducer does not deadlock with itself. This
can be expressed as a will-correctness property, that for any intermediate
state $U$, if $T_{xy}^{n}\mbox{ }|\mbox{ }tx^{n}\rightarrow^{*}U$
then $U\rightarrow^{*}ty^{n}$. This appears to require an induction
on all possible intermediate configurations $U$ for any $n$. Even
for a fixed small $n$, the state space $U$ can grow very large,
which suggests that automated state exploration tools should be useful.
Note also that an induction on the length of $\rightarrow^{*}$ is
problematic because of the reversible exchange rule: infinite sequences
of reductions exist in almost all systems. In a stochastic interpretation
of reduction, actual convergence can often be achieved (with measure
$1$), and this is another challenging property to prove. 

We now illustrate how to check a will-correctness property, for a
single copy of a transducer:

\subsubsection*{\label{pro:Tranducer-Will-Correctness}Proposition 4: $T_{xy}^{1}$
Will-Correctness }

\vspace{0in}

$T_{xy}^{1}\mbox{ }|\mbox{ }{}_{\ulcorner}x\rightarrow^{\forall}{}_{\ulcorner}y$.
$\quad$Moreover, ${}_{\ulcorner}y$ is the only reachable terminal
state.

\subsubsection*{Proof}

We show that if $T_{xy}^{1}\mbox{ }|\mbox{ }{}_{\ulcorner}x\rightarrow^{*}U$
then $U\rightarrow^{*}{}_{\ulcorner}y$. We enumerate all distinct
states $U$, up to algebraic equality, arising from $T_{xy}^{1}\mbox{ }|\mbox{ }tx$
by all possible traces, and then we check that each state can lead
to ${}_{\ulcorner}y$. Assume $x\neq y$; indentation means a branch
in the derivation:

01. $(\nu a)\mbox{ }\underline{_{\smile}x_{\urcorner}a_{\urcorner}a}\mbox{ }|\mbox{ }_{\ulcorner}a\mbox{ }|\mbox{ }\underline{x{}_{\ulcorner}y_{\ulcorner}a_{\smile}}\mbox{ }|\mbox{ }y_{\urcorner}\mbox{ }|\mbox{ }{}_{\ulcorner}x$

02. $\leftrightarrow$ $(\nu a)\mbox{ }\underline{{}_{\ulcorner}x_{\smile}a_{\urcorner}a}\mbox{ }|\mbox{ }_{\ulcorner}a\mbox{ }|\mbox{ }\underline{x{}_{\ulcorner}y_{\ulcorner}a_{\smile}}\mbox{ }|\mbox{ }y_{\urcorner}\mbox{ }|\mbox{ }x_{\urcorner}$

03. $\leftrightarrow$ $(\nu a)\mbox{ }\underline{{}_{\ulcorner}x_{\ulcorner}a_{\smile}a}\mbox{ }|\mbox{ }\underline{x{}_{\ulcorner}y_{\ulcorner}a_{\smile}}\mbox{ }|\mbox{ }y_{\urcorner}\mbox{ }|\mbox{ }x_{\urcorner}\mbox{ }|\mbox{ }a_{\urcorner}$

04. $\leftrightarrow$ $(\nu a)\mbox{ }\underline{{}_{\ulcorner}x_{\ulcorner}a_{\smile}a}\mbox{ }|\mbox{ }\underline{x{}_{\ulcorner}y_{\smile}a_{\urcorner}}\mbox{ }|\mbox{ }y_{\urcorner}\mbox{ }|\mbox{ }x_{\urcorner}\mbox{ }|\mbox{ }_{\ulcorner}a$

05. $\ \ \ \ \ \ \ \ \leftrightarrow$ $(\nu a)\mbox{ }\underline{{}_{\ulcorner}x_{\ulcorner}a_{\ulcorner}a}\mbox{ }|\mbox{ }\underline{x{}_{\ulcorner}y_{\smile}a_{\urcorner}}\mbox{ }|\mbox{ }y_{\urcorner}\mbox{ }|\mbox{ }x_{\urcorner}$

06. $\ \ \ \ \ \ \ \ \ \ \ \ \ \ \ \ \rightarrow$ $(\nu a)\mbox{ }\underline{x{}_{\ulcorner}y_{\smile}a_{\urcorner}}\mbox{ }|\mbox{ }y_{\urcorner}\mbox{ }|\mbox{ }x_{\urcorner}$

07. $\ \ \ \ \ \ \ \ \ \ \ \ \ \ \ \ \leftrightarrow$ $(\nu a)\mbox{ }\underline{x_{\smile}y_{\urcorner}a_{\urcorner}}\mbox{ }|\mbox{ }x_{\urcorner}\mbox{ }|\mbox{ }{}_{\ulcorner}y$

08. $\ \ \ \ \ \ \ \ \ \ \ \ \ \ \ \ \leftrightarrow$ $(\nu a)\mbox{ }\underline{x_{\urcorner}y_{\urcorner}a_{\urcorner}}\mbox{ }|\mbox{ }{}_{\ulcorner}y$

09. $\ \ \ \ \ \ \ \ \ \ \ \ \ \ \ \ \rightarrow$ ${}_{\ulcorner}y$

10. $\ \ \ \ \ \ \ \ \leftrightarrow(\nu a)\mbox{ }\underline{{}_{\ulcorner}x_{\ulcorner}a_{\ulcorner}a}\mbox{ }|\mbox{ }\underline{x_{\smile}y_{\urcorner}a_{\urcorner}}\mbox{ }|\mbox{ }x_{\urcorner}\mbox{ }|\mbox{ }{}_{\ulcorner}y$$\ \ \ \ \ \ \rightarrow$
07

11. $\ \ \ \ \ \ \ \ \leftrightarrow(\nu a)\mbox{ }\underline{{}_{\ulcorner}x_{\ulcorner}a_{\ulcorner}a}\mbox{ }|\underline{\mbox{ }x_{\urcorner}y_{\urcorner}a_{\urcorner}}\mbox{ }|\mbox{ }{}_{\ulcorner}y$$\ \ \ \ \ \ \rightarrow$
08

12. $\ \ \ \ \ \ \ \ \leftrightarrow(\nu a)\mbox{ }\underline{{}_{\ulcorner}x_{\ulcorner}a_{\ulcorner}a}\mbox{ }|\mbox{ }{}_{\ulcorner}y$$\ \ \ \ \ \ \rightarrow$
09

13. $\leftrightarrow(\nu a)\mbox{ }\underline{{}_{\ulcorner}x_{\ulcorner}a_{\smile}a}\mbox{ }|\underline{\mbox{ }x_{\smile}y_{\urcorner}a_{\urcorner}}\mbox{ }|\mbox{ }x_{\urcorner}\mbox{ }|\mbox{ }_{\ulcorner}a\mbox{ }|\mbox{ }{}_{\ulcorner}y$$\ \ \ \ \ \ \leftrightarrow$
10

14. $\leftrightarrow(\nu a)\mbox{ }\underline{{}_{\ulcorner}x_{\ulcorner}a_{\smile}a}\mbox{ }|\mbox{ }\underline{x_{\urcorner}y_{\urcorner}a_{\urcorner}}\mbox{ }|\mbox{ }_{\ulcorner}a\mbox{ }|\mbox{ }{}_{\ulcorner}y$$\ \ \ \ \ \ \leftrightarrow$
11

15. $\leftrightarrow(\nu a)\mbox{ }\underline{{}_{\ulcorner}x_{\ulcorner}a_{\smile}a}\mbox{ }|\mbox{ }_{\ulcorner}a\mbox{ }|\mbox{ }{}_{\ulcorner}y$$\ \ \ \ \ \ \leftrightarrow$
12

All other states (up to algebraic equality) can be reduced to these
states by well-mixing. We can then check that all these states have
a path to state $9$. The case for $x=y$ is similar: the state graphs
is the same because, as can be seen above, there is never both an
$x$ redex and a different $y$ redex in the same state, and when
two $x$ signals or cosignals can be chosen, it does not matter which
one is chosen, by well-mixing. \textbf{End proof.}

For transducer composition, the may-correctness property $T_{xy}^{n}\mbox{ }|\mbox{ }T_{yz}^{n}\mbox{ }|\mbox{ }{}_{\ulcorner}x^{n}\rightarrow^{*}{}_{\ulcorner}z^{n}$
follows simply from Proposition 1, but even just the will-correctness
property $T_{xy}^{1}$ $\mbox{ }|\mbox{ }$ $T_{yz}^{1}$ $\mbox{ }|\mbox{ }$
${}_{\ulcorner}x$ $\rightarrow^{\forall}$ ${}_{\ulcorner}z$ (including
$x=z$ and $y=z$ and $x=y=z$) does not follow from Proposition 4,
and requires the analysis of a product state space. For example, $T_{xy}^{1}\mbox{ }|\mbox{ }T_{yx}^{1}$
can absorb the inputs ${}_{\ulcorner}x\mbox{ }|\mbox{ }{}_{\ulcorner}y$
sequentially (converting ${}_{\ulcorner}x$ to a second ${}_{\ulcorner}y$
and then ${}_{\ulcorner}y$ to ${}_{\ulcorner}x$) or in parallel
(each transducer starting to process an input before producing an
output). In fact, consider the following transducer that uses a public
\textquoteleft{}$a$\textquoteright{} domain instead of a private
one:

\medskip{}

$T_{xay}=\underline{{}_{\smile}x_{\urcorner}a_{\urcorner}a}\mbox{ }|\mbox{ }_{\ulcorner}a\mbox{ }|\mbox{ }\underline{x{}_{\ulcorner}y_{\ulcorner}a_{\smile}}\mbox{ }|\mbox{ }y_{\urcorner}$

\medskip{}

$T_{xay}$ by itself satisfies may and will-correctness as shown above
for $T_{xy}^{1}$, and so does $T_{yax}$. But the two together do
not satisfy the will-correctness property of just producing ${}_{\ulcorner}x$
on input ${}_{\ulcorner}x$, because the following \textquoteleft{}crosstalk\textquoteright{}
derivation is possible, where in the third step $a_{\urcorner}$ goes
to the \textquoteleft{}wrong\textquoteright{} gate:

\medskip{}

$T_{xay}\mbox{ }|\mbox{ }T_{yax}\mbox{ }|\mbox{ }{}_{\ulcorner}x$

$=\underline{_{\smile}x_{\urcorner}a_{\urcorner}a}\mbox{ }|\mbox{ }_{\ulcorner}a\mbox{ }|\mbox{ }\underline{x{}_{\ulcorner}y_{\ulcorner}a_{\smile}}\mbox{ }|\mbox{ }y_{\urcorner}\mbox{ }|\mbox{ }\underline{_{\smile}y_{\urcorner}a_{\urcorner}a}\mbox{ }|\mbox{ }_{\ulcorner}a\mbox{ }|\mbox{ }\underline{y{}_{\ulcorner}x_{\ulcorner}a_{\smile}}\mbox{ }|\mbox{ }x_{\urcorner}\mbox{ }|\mbox{ }{}_{\ulcorner}x$

$\leftrightarrow\underline{{}_{\ulcorner}x_{\smile}a_{\urcorner}a}\mbox{ }|\mbox{ }_{\ulcorner}a\mbox{ }|\mbox{ }\underline{x{}_{\ulcorner}y_{\ulcorner}a_{\smile}}\mbox{ }|\mbox{ }y_{\urcorner}\mbox{ }|\mbox{ }\underline{_{\smile}y_{\urcorner}a_{\urcorner}a}\mbox{ }|\mbox{ }_{\ulcorner}a\mbox{ }|\mbox{ }y{}_{\ulcorner}x_{\ulcorner}a_{\smile}\mbox{ }|\mbox{ }x_{\urcorner}\mbox{ }|\mbox{ }x_{\urcorner}$

$\leftrightarrow\underline{{}_{\ulcorner}x_{\ulcorner}a_{\smile}a}\mbox{ }|\mbox{ }\underline{x{}_{\ulcorner}y_{\ulcorner}a_{\smile}}\mbox{ }|\mbox{ }y_{\urcorner}\mbox{ }|\mbox{ }\underline{_{\smile}y_{\urcorner}a_{\urcorner}a}\mbox{ }|\mbox{ }_{\ulcorner}a\mbox{ }|\mbox{ }\underline{y{}_{\ulcorner}x_{\ulcorner}a_{\smile}}\mbox{ }|\mbox{ }x_{\urcorner}\mbox{ }|\mbox{ }x_{\urcorner}\mbox{ }|\mbox{ }a_{\urcorner}$

$\leftrightarrow\underline{{}_{\ulcorner}x_{\ulcorner}a_{\smile}a}\mbox{ }|\mbox{ }\underline{x{}_{\ulcorner}y_{\ulcorner}a_{\smile}}\mbox{ }|\mbox{ }y_{\urcorner}\mbox{ }|\mbox{ }\underline{_{\smile}y_{\urcorner}a_{\urcorner}a}\mbox{ }|\mbox{ }_{\ulcorner}a\mbox{ }|\mbox{ }\underline{y{}_{\ulcorner}x_{\smile}a_{\urcorner}}\mbox{ }|\mbox{ }x_{\urcorner}\mbox{ }|\mbox{ }x_{\urcorner}\mbox{ }|\mbox{ }_{\ulcorner}a$

$\rightarrow\underline{{}_{\ulcorner}x_{\ulcorner}a_{\ulcorner}a}\mbox{ }|\mbox{ }\underline{x{}_{\ulcorner}y_{\ulcorner}a_{\smile}}\mbox{ }|\mbox{ }y_{\urcorner}\mbox{ }|\mbox{ }\underline{_{\smile}y_{\urcorner}a_{\urcorner}a}\mbox{ }|\mbox{ }_{\ulcorner}a\mbox{ }|\mbox{ }\underline{y{}_{\ulcorner}x_{\smile}a_{\urcorner}}\mbox{ }|\mbox{ }x_{\urcorner}\mbox{ }|\mbox{ }x_{\urcorner}$

$\rightarrow\underline{x{}_{\ulcorner}y_{\ulcorner}a_{\smile}}\mbox{ }|\mbox{ }y_{\urcorner}\mbox{ }|\mbox{ }\underline{_{\smile}y_{\urcorner}a_{\urcorner}a}\mbox{ }|\mbox{ }_{\ulcorner}a\mbox{ }|\mbox{ }\underline{y{}_{\ulcorner}x_{\smile}a_{\urcorner}}\mbox{ }|\mbox{ }x_{\urcorner}\mbox{ }|\mbox{ }x_{\urcorner}$

$\leftrightarrow\underline{x{}_{\ulcorner}y_{\ulcorner}a_{\smile}}\mbox{ }|\mbox{ }y_{\urcorner}\mbox{ }|\mbox{ }\underline{_{\smile}y_{\urcorner}a_{\urcorner}a}\mbox{ }|\mbox{ }_{\ulcorner}a\mbox{ }|\mbox{ }\underline{y_{\smile}x_{\urcorner}a_{\urcorner}}\mbox{ }|\mbox{ }x_{\urcorner}\mbox{ }|\mbox{ }{}_{\ulcorner}x$

$\rightarrow\underline{x{}_{\ulcorner}y_{\ulcorner}a_{\smile}}\mbox{ }|\mbox{ }\underline{_{\smile}y_{\urcorner}a_{\urcorner}a}\mbox{ }|\mbox{ }_{\ulcorner}a\mbox{ }|\mbox{ }\underline{y_{\urcorner}x_{\urcorner}a_{\urcorner}}\mbox{ }|\mbox{ }x_{\urcorner}\mbox{ }|\mbox{ }{}_{\ulcorner}x$

$\rightarrow\underline{x{}_{\ulcorner}y_{\ulcorner}a_{\smile}}\mbox{ }|\mbox{ }\underline{_{\smile}y_{\urcorner}a_{\urcorner}a}\mbox{ }|\mbox{ }_{\ulcorner}a\mbox{ }|\mbox{ }x_{\urcorner}\mbox{ }|\mbox{ }{}_{\ulcorner}x$

\medskip{}

The last state is final (no further progress can be made), and is
not just the expected ${}_{\ulcorner}x$ (which can be obtained by
a different derivation). Moreover, no ${}_{\ulcorner}y$ is ever produced.
The system is deadlocked in a state where the output ${}_{\ulcorner}x$
has been produced, but many other active components have been left
to interfere with future operation. However, that last state, if supplied
with an additional ${}_{\ulcorner}y$, then unblocks and reduces just
to ${}_{\ulcorner}y\mbox{ }|\mbox{ }{}_{\ulcorner}x$. Hence, although
$T_{xay}\mbox{ }|\mbox{ }T_{yax}\mbox{ }|\mbox{ }{}_{\ulcorner}x$
$\nrightarrow^{\forall}$ ${}_{\ulcorner}x$, we have that $T_{xay}\mbox{ }|\mbox{ }T_{yax}\mbox{ }|\mbox{ }{}_{\ulcorner}x\mbox{ }|\mbox{ }{}_{\ulcorner}y$
$\rightarrow^{\forall}$ ${}_{\ulcorner}x\mbox{ }|\mbox{ }{}_{\ulcorner}y$.
That means that a large population of such gates in practice does
not deadlock easily over an input population of ${}_{\ulcorner}x$:
each pair of stuck gates can be unblocked by another gate correctly
producing a ${}_{\ulcorner}y$, and it is very unlikely that a large
fraction of gates ends up being blocked. This can be seen in stochastic
simulations of large populations, and also in Ordinary Differential
Equation simulations with unit concentration of $T_{xay}\mbox{ }|\mbox{ }T_{yax}$,
where the concentration of the residual $_{\ulcorner}a$ tends asymptotically
to zero. Hence, another interesting property of these system is that,
even though small populations may deadlock, large populations may
converge to an almost-correct solution with high probability.

\section{Testing }

Gate and circuits designs have been tested with the DSD tool \cite{Phillips. A Programming Language for Composable DNA Circuits}.
We give a simple example here, testing a combination of two fork and
four join gates in the following configuration, where $yv$, $yw$,
$zv$, $zw$ are four output domains (i.e., $yv$ does not mean $y.v$
in this section).

\medskip{}

\begin{minipage}[t]{1\columnwidth}%
$F_{x,y,z}^{n}$$\mbox{ }|\mbox{ }$$\ \ \ \ \ \ \ \ \ \ \ \ \ \ \ \ \ \ \ \ \ \ \ \ \ \ \ \ \ \ \ \ \ $\includegraphics[bb=0bp 176bp 183bp 176bp,scale=0.4]{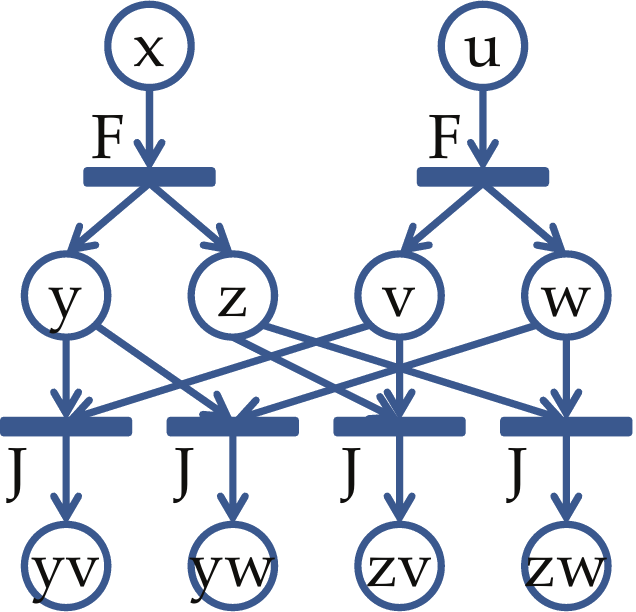}

$F_{u,v,w}^{n}$$\mbox{ }|\mbox{ }$

$J_{y,v,yv}^{n}$$\mbox{ }|\mbox{ }$

$J_{y,w,yw}^{n}$$\mbox{ }|\mbox{ }$

$J_{z,v,zv}^{n}$$\mbox{ }|\mbox{ }$

$J_{z,w,zw}^{n}$$\mbox{ }|\mbox{ }$

${}_{\ulcorner}x^{m}\mbox{ }|\mbox{ }{}_{\ulcorner}u^{m}$ (input,
$m\leq n$)%
\end{minipage}

\medskip{}

\begin{figure}
\begin{centering}
\includegraphics[scale=0.3]{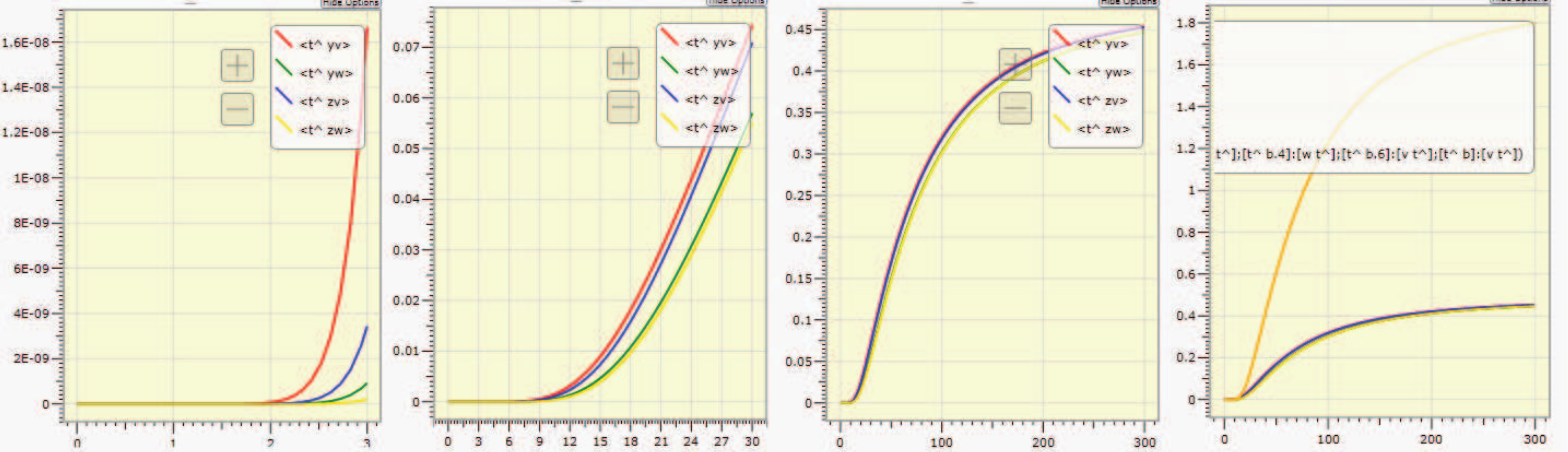}
\par\end{centering}

\caption{\label{fig:Testing a circuit}Testing a fork/join circuit.}

\end{figure}

Since fork and join gates accept inputs and produce outputs in a specific
order, one should not expect identical rates of production of $yv$,$yw$,$zv$,$zw$.
(If desired, one can mix populations of symmetric gates, to achieve
symmetric behavior.) In Figure \ref{fig:Testing a circuit} we see
an Ordinary Differential Equations simulation with unit rates for
toehold binding and unbinding, and with concentrations of $1.0$ for
the input signals and $10.0$ for the gates; hence $10\%$ of each
gates is consumed during the computation. The system has a total of
$54$ single strand species, $108$ double strand species, and $172$
reactions, and therefore $162$ ODEs.%
At time $3$ (left), $yv$ is ahead out of the gates, with $zw$ trailing
last. At time $30$ (middle left) $yv$ and $yw$ are closer, and
$zv$ and $zw$ are closer. At time $300$ (middle right) the computation
has reached $90\%$ completion with similar output quantities approaching
the expected $0.5$ concentration. The higher curve of the fourth
graph shows the total accumulation of the four $\underline{_{\ulcorner}D^{\lyxmathsym{\dag}}D\lyxmathsym{\textquoteright}_{\urcorner}}$
garbage species for the join gates, indicating that all the gates
are being converted to waste. One can further examine the trajectories
of all the species in the system to check that no deadlock occurs,
and that all the structures are turned to output or to waste.

\section{Conclusions }

We have shown how to implement fork and join gates via simple two-domain
structures, and how to implement them in a \textquoteleft{}clean\textquoteright{}
way that automatically removes all active garbage. In essence, we
have given an implementation of the higher-level strand algebra of
\cite{Cardelli. Strand Algebras for DNA Computing}. But is this implementation
correct? We have provided a formal framework where we can perform
calculations and study such questions, and we have discussed some
simple correctness definitions and some complex behavioral properties.
A formal proof of absence of gate interference under all possible
combinations and numbers of gates and inputs will require an extensive
amount of case analysis, which likely needs to be automated, as well
as the identification of appropriate invariants. Alternatively, one
may gain confidence in the designs by simulation testing.

\subsection*{Acknowledgments}

Figures were prepared with the DSD tool \cite{Phillips. A Programming Language for Composable DNA Circuits}.
I would like to thank the members of the Molecular Programming Project
at Caltech and U.Washington for many tutorials and discussions.

\section{Appendix }

\subsection{May-Correctness of binary Fork and Join gates }

\subsubsection*{Proposition 2: $F_{xyz}^{n}$ May-Correctness}

\vspace{0in}

Let $F_{xyz}^{n}=(\nu a)((\underline{_{\smile}x_{\urcorner}a_{\urcorner}a}\mbox{ }|\mbox{ }_{\ulcorner}a\mbox{ }|\mbox{ }\underline{x_{\ulcorner}z_{\ulcorner}y_{\ulcorner}a_{\smile}}\mbox{ }|\mbox{ }z_{\urcorner}\mbox{ }|\mbox{ }y_{\urcorner})^{n})$,

then $F_{xyz}^{n}\mbox{ }|\mbox{ }_{\ulcorner}x^{n}\rightarrow^{*}{}_{\ulcorner}y^{n}\mbox{ }|\mbox{ }_{\ulcorner}z^{n}$.

\subsubsection*{Proof}

Let $F{}_{xayz}=\underline{{}_{\smile}x_{\urcorner}a_{\urcorner}a}\mbox{ }|\mbox{ }{}_{\ulcorner}a\mbox{ }|\mbox{ }\underline{x{}_{\ulcorner}z{}_{\ulcorner}y{}_{\ulcorner}a_{\smile}}\mbox{ }|\mbox{ }z_{\urcorner}\mbox{ }|\mbox{ }y_{\urcorner}$
for $a\neq x,y,z$, so that$F_{xyz}^{n}=$ $(\nu a)((F{}_{xayz})^{n})$.
We first show that $F{}_{xayz}\mbox{ }|\mbox{ }{}_{\ulcorner}x\rightarrow^{*}{}_{\ulcorner}y\mbox{ }|\mbox{ }{}_{\ulcorner}z$. 

$F{}_{xayz}\mbox{ }|\mbox{ }{}_{\ulcorner}x$

$=\underline{{}_{\smile}x_{\urcorner}a_{\urcorner}}a\mbox{ }|\mbox{ }{}_{\ulcorner}a\mbox{ }|\underline{\mbox{ }x{}_{\ulcorner}z{}_{\ulcorner}y{}_{\ulcorner}a_{\smile}}\mbox{ }|\mbox{ }z_{\urcorner}\mbox{ }|\mbox{ }y_{\urcorner}\mbox{ }|\mbox{ }x_{\urcorner}$

$\leftrightarrow\underline{{}_{\ulcorner}x_{\smile}a_{\urcorner}a}\mbox{ }|\mbox{ }{}_{\ulcorner}a\mbox{ }|\mbox{ }\underline{x{}_{\ulcorner}z{}_{\ulcorner}y{}_{\ulcorner}a_{\smile}}\mbox{ }|\mbox{ }z_{\urcorner}\mbox{ }|\mbox{ }y_{\urcorner}\mbox{ }|\mbox{ }x_{\urcorner}$

$\leftrightarrow\underline{{}_{\ulcorner}x{}_{\ulcorner}a_{\smile}a}\mbox{ }|\mbox{ }\underline{x{}_{\ulcorner}z{}_{\ulcorner}y{}_{\ulcorner}a_{\smile}}\mbox{ }|\mbox{ }z_{\urcorner}\mbox{ }|\mbox{ }y_{\urcorner}\mbox{ }|\mbox{ }x_{\urcorner}\mbox{ }|\mbox{ }a_{\urcorner}$

$\leftrightarrow\underline{{}_{\ulcorner}x{}_{\ulcorner}a_{\smile}a}\mbox{ }|\mbox{ }\underline{x{}_{\ulcorner}z{}_{\ulcorner}y_{\smile}a_{\urcorner}}\mbox{ }|\mbox{ }z_{\urcorner}\mbox{ }|\mbox{ }y_{\urcorner}\mbox{ }|\mbox{ }x_{\urcorner}\mbox{ }|\mbox{ }{}_{\ulcorner}a$

$\rightarrow\underline{{}_{\ulcorner}x{}_{\ulcorner}a{}_{\ulcorner}a}\mbox{ }|\mbox{ }\underline{x{}_{\ulcorner}z{}_{\ulcorner}y_{\smile}a_{\urcorner}}\mbox{ }|\mbox{ }z_{\urcorner}\mbox{ }|\mbox{ }y_{\urcorner}\mbox{ }|\mbox{ }x_{\urcorner}$

$\rightarrow\underline{x{}_{\ulcorner}z{}_{\ulcorner}y_{\smile}a_{\urcorner}}\mbox{ }|\mbox{ }z_{\urcorner}\mbox{ }|\mbox{ }y_{\urcorner}\mbox{ }|\mbox{ }x_{\urcorner}$

$\leftrightarrow\underline{x{}_{\ulcorner}z_{\smile}y_{\urcorner}a_{\urcorner}}\mbox{ }|\mbox{ }z_{\urcorner}\mbox{ }|\mbox{ }x_{\urcorner}\mbox{ }|\mbox{ }{}_{\ulcorner}y$

$\leftrightarrow\underline{x_{\smile}z_{\urcorner}y_{\urcorner}a_{\urcorner}}\mbox{ }|\mbox{ }x_{\urcorner}\mbox{ }|\mbox{ }{}_{\ulcorner}y\mbox{ }|\mbox{ }{}_{\ulcorner}z$

$\rightarrow\underline{x_{\urcorner}z_{\urcorner}y_{\urcorner}a_{\urcorner}}\mbox{ }|\mbox{ }{}_{\ulcorner}y\mbox{ }|\mbox{ }{}_{\ulcorner}z$

$\rightarrow\mbox{ }|\mbox{ }{}_{\ulcorner}y\mbox{ }|\mbox{ }{}_{\ulcorner}z$

Hence $(F{}_{xayz}\mbox{ }|\mbox{ }{}_{\ulcorner}x)^{n}\rightarrow^{*}({}_{\ulcorner}y\mbox{ }|\mbox{ }{}_{\ulcorner}z)^{n}$
by induction, $(F{}_{xayz})^{n}\mbox{ }|\mbox{ }{}_{\ulcorner}x^{n}$
$\rightarrow$ ${}_{\ulcorner}y^{n}\mbox{ }|\mbox{ }{}_{\ulcorner}z^{n}$
by associativity, $(\nu a)((F{}_{xayz})^{n}\mbox{ }|\mbox{ }{}_{\ulcorner}x_{n})$
$\rightarrow^{*}(\nu a)({}_{\ulcorner}y^{n}\mbox{ }|\mbox{ }{}_{\ulcorner}z^{n})$
by isolation, and $F_{xyz}^{n}$ $\mbox{ }|\mbox{ }$ ${}_{\ulcorner}x^{n}$
$\rightarrow^{*}{}_{\ulcorner}y^{n}\mbox{ }|\mbox{ }{}_{\ulcorner}z^{n}$
by $\nu$-equivalence and by $F_{xyz}^{n}$ definition. \textbf{End
proof.}

\medskip{}

\subsubsection*{Proposition 3:\textmd{ $J_{xyz}^{n}$} May-Correctness \vspace{0in}
}

Let $J_{xyz}^{n}=(\nu a)(\nu b)((\underline{_{\smile}x_{\urcorner}y_{\urcorner}a_{\urcorner}a}\mbox{ }|\mbox{ }_{\ulcorner}a\mbox{ }|\mbox{ }\underline{x_{\ulcorner}b_{\ulcorner}z_{\ulcorner}a_{\smile}}\mbox{ }|\mbox{ }b_{\urcorner}\mbox{ }|\mbox{ }z_{\urcorner}\mbox{ }|\mbox{ }\underline{_{\smile}b^{\lyxmathsym{\dag}}y_{\smile}})^{n})$,

then $J_{xyz}^{n}\mbox{ }|\mbox{ }_{\ulcorner}x^{n}\mbox{ }|\mbox{ }_{\ulcorner}y^{n}\rightarrow^{*}{}_{\ulcorner}z^{n}$.

\subsubsection*{Proof}

Let $J{}_{xyaz}=\underline{{}_{\smile}x_{\urcorner}y_{\urcorner}a_{\urcorner}a}\mbox{ }|\mbox{ }{}_{\ulcorner}a\mbox{ }|\underline{\mbox{ }x{}_{\ulcorner}b{}_{\ulcorner}z{}_{\ulcorner}a_{\smile}}\mbox{ }|\mbox{ }b_{\urcorner}\mbox{ }|\mbox{ }z_{\urcorner}\mbox{ }|\mbox{ }\underline{{}_{\smile}b^{\lyxmathsym{\dag}}y_{\smile}}$
for $a\neq x,y,z$, so that $J_{xyz}^{n}=(\nu a)((J{}_{xyaz})^{n})$.
We first show that $J{}_{xyaz}\mbox{ }|\mbox{ }{}_{\ulcorner}x\mbox{ }|\mbox{ }{}_{\ulcorner}y\rightarrow^{*}{}_{\ulcorner}z$. 

$J{}_{xyaz}\mbox{ }|\mbox{ }{}_{\ulcorner}x\mbox{ }|\mbox{ }{}_{\ulcorner}y$

$=\underline{{}_{\smile}x_{\urcorner}y_{\urcorner}a_{\urcorner}a}\mbox{ }|\mbox{ }{}_{\ulcorner}a\mbox{ }|\mbox{ }\underline{x{}_{\ulcorner}b{}_{\ulcorner}z{}_{\ulcorner}a_{\smile}}\mbox{ }|\mbox{ }b_{\urcorner}\mbox{ }|\mbox{ }z_{\urcorner}\mbox{ }|\mbox{ }\underline{{}_{\smile}b^{\lyxmathsym{\dag}}y_{\smile}}\mbox{ }|\mbox{ }{}_{\ulcorner}x\mbox{ }|\mbox{ }{}_{\ulcorner}y$

$\leftrightarrow\underline{{}_{\ulcorner}x_{\smile}y_{\urcorner}a_{\urcorner}a}\mbox{ }|\mbox{ }{}_{\ulcorner}a\mbox{ }|\mbox{ }\underline{x{}_{\ulcorner}b{}_{\ulcorner}z{}_{\ulcorner}a_{\smile}}\mbox{ }|\mbox{ }b_{\urcorner}\mbox{ }|\mbox{ }z_{\urcorner}\mbox{ }|\mbox{ }\underline{{}_{\smile}b^{\lyxmathsym{\dag}}y_{\smile}}\mbox{ }|\mbox{ }{}_{\ulcorner}y\mbox{ }|\mbox{ }x_{\urcorner}$

$\leftrightarrow\underline{{}_{\ulcorner}x{}_{\ulcorner}y_{\smile}a_{\urcorner}a}\mbox{ }|\mbox{ }{}_{\ulcorner}a\mbox{ }|\mbox{ }\underline{x{}_{\ulcorner}b{}_{\ulcorner}z{}_{\ulcorner}a_{\smile}}\mbox{ }|\mbox{ }b_{\urcorner}\mbox{ }|\mbox{ }z_{\urcorner}\mbox{ }|\mbox{ }\underline{{}_{\smile}b^{\lyxmathsym{\dag}}y_{\smile}}\mbox{ }|\mbox{ }x_{\urcorner}\mbox{ }|\mbox{ }y_{\urcorner}$

$\leftrightarrow\underline{{}_{\ulcorner}x{}_{\ulcorner}y{}_{\ulcorner}a_{\smile}a}\mbox{ }|\mbox{ }\underline{x{}_{\ulcorner}b{}_{\ulcorner}z{}_{\ulcorner}a_{\smile}}\mbox{ }|\mbox{ }b_{\urcorner}\mbox{ }|\mbox{ }z_{\urcorner}\mbox{ }|\mbox{ }\underline{{}_{\smile}b^{\text{\dag}}y_{\smile}}\mbox{ }|\mbox{ }x_{\urcorner}\mbox{ }|\mbox{ }y_{\urcorner}\mbox{ }|\mbox{ }a_{\urcorner}$

$\leftrightarrow\underline{{}_{\ulcorner}x{}_{\ulcorner}y{}_{\ulcorner}a_{\smile}a}\mbox{ }|\mbox{ }\underline{x{}_{\ulcorner}b{}_{\ulcorner}z_{\smile}a_{\urcorner}}\mbox{ }|\mbox{ }b_{\urcorner}\mbox{ }|\mbox{ }z_{\urcorner}\mbox{ }|\mbox{ }\underline{{}_{\smile}b^{\text{\dag}}y_{\smile}}\mbox{ }|\mbox{ }x_{\urcorner}\mbox{ }|\mbox{ }y_{\urcorner}\mbox{ }|\mbox{ }{}_{\ulcorner}a$

$\rightarrow\underline{{}_{\ulcorner}x{}_{\ulcorner}y{}_{\ulcorner}a{}_{\ulcorner}a}\mbox{ }|\mbox{ }\underline{x{}_{\ulcorner}b{}_{\ulcorner}z_{\smile}a_{\urcorner}}\mbox{ }|\mbox{ }b_{\urcorner}\mbox{ }|\mbox{ }z_{\urcorner}\mbox{ }|\mbox{ }\underline{{}_{\smile}b^{\text{\dag}}y_{\smile}}\mbox{ }|\mbox{ }x_{\urcorner}\mbox{ }|\mbox{ }y_{\urcorner}$

$\rightarrow\underline{x{}_{\ulcorner}b{}_{\ulcorner}z_{\smile}a_{\urcorner}}\mbox{ }|\mbox{ }b_{\urcorner}\mbox{ }|\mbox{ }z_{\urcorner}\mbox{ }|\mbox{ }\underline{{}_{\smile}b^{\text{\dag}}y_{\smile}}\mbox{ }|\mbox{ }x_{\urcorner}\mbox{ }|\mbox{ }y_{\urcorner}$

$\leftrightarrow\underline{x{}_{\ulcorner}b_{\smile}z_{\urcorner}a_{\urcorner}}\mbox{ }|\mbox{ }b_{\urcorner}\mbox{ }|\mbox{ }\underline{{}_{\smile}b^{\text{\dag}}y_{\smile}}\mbox{ }|\mbox{ }x_{\urcorner}\mbox{ }|\mbox{ }y_{\urcorner}\mbox{ }|\mbox{ }{}_{\ulcorner}z$

$\leftrightarrow\underline{x_{\smile}b_{\urcorner}z_{\urcorner}a_{\urcorner}}\mbox{ }|\mbox{ }\underline{{}_{\smile}b^{\text{\dag}}y_{\smile}}\mbox{ }|\mbox{ }x_{\urcorner}\mbox{ }|\mbox{ }y_{\urcorner}\mbox{ }|\mbox{ }{}_{\ulcorner}z\mbox{ }|\mbox{ }{}_{\ulcorner}b$

$\rightarrow\underline{x_{\urcorner}b_{\urcorner}z_{\urcorner}a_{\urcorner}}\mbox{ }|\mbox{ }\underline{{}_{\smile}b^{\text{\dag}}y_{\smile}}\mbox{ }|\mbox{ }y_{\urcorner}\mbox{ }|\mbox{ }{}_{\ulcorner}z\mbox{ }|\mbox{ }{}_{\ulcorner}b$

$\rightarrow\underline{{}_{\smile}b^{\text{\dag}}y_{\smile}}\mbox{ }|\mbox{ }y_{\urcorner}\mbox{ }|\mbox{ }{}_{\ulcorner}z\mbox{ }|\mbox{ }{}_{\ulcorner}b$

$\rightarrow{}_{\ulcorner}z$

Hence $(J{}_{xyaz}\mbox{ }|\mbox{ }{}_{\ulcorner}x\mbox{ }|\mbox{ }{}_{\ulcorner}y)^{n}\rightarrow^{*}{}_{\ulcorner}z^{n}$
by induction, $(J{}_{xyaz})^{n}\mbox{ }|\mbox{ }{}_{\ulcorner}x^{n}\mbox{ }|\mbox{ }{}_{\ulcorner}y^{n}\rightarrow^{*}{}_{\ulcorner}z^{n}$
by associativity, $(\nu a)((J{}_{xyaz})^{n}\mbox{ }|\mbox{ }{}_{\ulcorner}x^{n}\mbox{ }|\mbox{ }{}_{\ulcorner}y^{n})\rightarrow^{*}(\nu a){}_{\ulcorner}z^{n}$
by isolation, and $J_{xyz}^{n}$ $\mbox{ }|\mbox{ }$ ${}_{\ulcorner}x^{n}\mbox{ }|\mbox{ }{}_{\ulcorner}y^{n}\rightarrow^{*}{}_{\ulcorner}z^{n}$
by $\nu$-equivalence and by $J_{xyz}^{n}$ definition. \textbf{End
proof.}

\subsection{DSD Script for Figure \ref{fig:Testing a circuit} }

This script can be run from a browser in DSD \cite{Phillips. A Programming Language for Composable DNA Circuits}
using \textquoteleft{}deterministic\textquoteright{} simulation. 

\textit{http://research.microsoft.com/en-us/projects/dna/default.aspx}

\medskip{}

\texttt{\scriptsize directive sample 300.0 1000 }{\scriptsize \par}

\texttt{\scriptsize directive plot <t\textasciicircum{} yv>; <t\textasciicircum{}
yw>; <t\textasciicircum{} zv>; <t\textasciicircum{} zw>; sum({[}t\textasciicircum{}
\_{]}:{[}\_ t\textasciicircum{}{]}) }{\scriptsize \par}

\texttt{\scriptsize new t@1.0,1.0}{\scriptsize \par}

\texttt{\scriptsize \vspace{0.01in}
}{\scriptsize \par}

\texttt{\scriptsize def F(N, x, y, z) = }{\scriptsize \par}

\texttt{\scriptsize new a }{\scriptsize \par}

\texttt{\scriptsize ( N{*} <t\textasciicircum{} a> }{\scriptsize \par}

\texttt{\scriptsize | N{*} <y t\textasciicircum{}> }{\scriptsize \par}

\texttt{\scriptsize | N{*} <z t\textasciicircum{}> }{\scriptsize \par}

\texttt{\scriptsize | N{*} t\textasciicircum{}:{[}x t\textasciicircum{}{]}:{[}a
t\textasciicircum{}{]}:{[}a{]} }{\scriptsize \par}

\texttt{\scriptsize | N{*} {[}x{]}:{[}t\textasciicircum{} z{]}:{[}t\textasciicircum{}
y{]}:{[}t\textasciicircum{} a{]}:t\textasciicircum{} )}{\scriptsize \par}

\texttt{\scriptsize \vspace{0.01in}
}{\scriptsize \par}

\texttt{\scriptsize def J(N, x, y, z) = }{\scriptsize \par}

\texttt{\scriptsize new a new b }{\scriptsize \par}

\texttt{\scriptsize ( N{*} <t\textasciicircum{} a> }{\scriptsize \par}

\texttt{\scriptsize | N{*} <b t\textasciicircum{}> }{\scriptsize \par}

\texttt{\scriptsize | N{*} <z t\textasciicircum{}> }{\scriptsize \par}

\texttt{\scriptsize | N{*} t\textasciicircum{}:{[}x t\textasciicircum{}{]}:{[}y
t\textasciicircum{}{]}:{[}a t\textasciicircum{}{]}:{[}a{]} }{\scriptsize \par}

\texttt{\scriptsize | N{*} {[}x{]}:{[}t\textasciicircum{} b{]}:{[}t\textasciicircum{}
z{]}:{[}t\textasciicircum{} a{]}:t\textasciicircum{} }{\scriptsize \par}

\texttt{\scriptsize | N{*} t\textasciicircum{}:{[}b y{]}:t\textasciicircum{}
)}{\scriptsize \par}

\texttt{\scriptsize \vspace{0.01in}
}{\scriptsize \par}

\texttt{\scriptsize ( F(10, x, y, z) }{\scriptsize \par}

\texttt{\scriptsize | F(10, u, v, w) }{\scriptsize \par}

\texttt{\scriptsize | J(10, y, v, yv) }{\scriptsize \par}

\texttt{\scriptsize | J(10, y, w, yw) }{\scriptsize \par}

\texttt{\scriptsize | J(10, z, v, zv) }{\scriptsize \par}

\texttt{\scriptsize | J(10, z, w, zw) }{\scriptsize \par}

\texttt{\scriptsize | 1 {*} <t\textasciicircum{} x> }{\scriptsize \par}

\texttt{\scriptsize | 1 {*} <t\textasciicircum{} u> ) }
\end{document}